\documentclass[aps, prl, floatfix,
  reprint, 
  superscriptaddress, 
  longbibliography
]{revtex4-2} 
\usepackage{graphicx, tikz, color} 
\graphicspath{ {./Figure/} }
\usepackage{physics, amsmath, amssymb} 
\usepackage{mathtools} 
\usepackage{natbib}
\usepackage[pdftex]{hyperref}
\usepackage{hypernat}
\hypersetup{
	colorlinks=false,
	bookmarksnumbered=true,
	pdfborder={0 0 0},
	bookmarkstype=toc,
	pdfauthor={Shoki Sugimoto, Ryusuke Hamazaki, and Masahito Ueda},
	pdftitle={TestOfETH}
}

\begin{document}
\title{
  Test of Eigenstate Thermalization Hypothesis Based on Local Random Matrix Theory
}
\author{Shoki Sugimoto}
  \affiliation{Department of Physics, University of Tokyo, 7-3-1 Hongo, Bunkyo-ku, Tokyo 113-0033, Japan}
\author{Ryusuke Hamazaki}
  \affiliation{Department of Physics, University of Tokyo, 7-3-1 Hongo, Bunkyo-ku, Tokyo 113-0033, Japan}
  \affiliation{Nonequilibrium Quantum Statistical Mechanics RIKEN Hakubi Research Team, RIKEN Cluster for Pioneering Research (CPR), RIKEN iTHEMS, Wako, Saitama 351-0198, Japan}
\author{Masahito Ueda}
  \affiliation{Department of Physics, University of Tokyo, 7-3-1 Hongo, Bunkyo-ku, Tokyo 113-0033, Japan}
  \affiliation{RIKEN Center for Emergent Matter Science (CEMS), Wako 351-0198, Japan}
\begin{abstract}
We verify that the eigenstate thermalization hypothesis (ETH) holds universally for locally interacting quantum many-body systems.
Introducing random-matrix ensembles with interactions, we numerically obtain a distribution of maximum fluctuations of eigenstate expectation values for different realizations of the interactions.
This distribution, which cannot be obtained from the conventional random matrix theory involving nonlocal correlations, demonstrates that an overwhelming majority of pairs of local Hamiltonians and observables satisfy the ETH with exponentially small fluctuations.
The ergodicity of our random matrix ensembles breaks down due to locality.
\end{abstract}
\maketitle

\paragraph{Introduction.}
Deriving statistical mechanics from unitary dynamics of isolated quantum systems has been a holy grail since von Neumann's seminal work~\cite{neumann1929beweis}. 
Last two decades have witnessed a resurgence of interest in this problem~\cite{polkovnikov2011colloquium,eisert2015quantum, gogolin2016equilibration, d2016quantum, mori2018thermalization} owing to the state-of-the-art experiments which utilize ultracold atoms~\cite{kinoshita2006quantum, trotzky2012probing, gring2012relaxation, langen2013local, schreiber2015observation, kaufman2016quantum, choi2016exploring} and ions~\cite{clos2016time, smith2016many} for artificial quantum simulators.

The eigenstate thermalization hypothesis (ETH) is widely accepted as the main scenario for thermalization of isolated quantum systems~\cite{deutsch1991quantum, srednicki1994chaos, rigol2008thermalization}. 
The ETH states that every energy eigenstate is thermal and ensures that any initial state relaxes to thermal equilibrium.
Despite considerable efforts~\cite{
biroli2010effect, santos2010localization, steinigeweg2013eigenstate, beugeling2014finite, de2015necessity, santos2010onset, goldstein2010long, goldstein2010approach, goldstein2010normal, ikeda2011eigenstate, ikeda2013finite, kim2014testing, steinigeweg2014pushing, reimann2015eigenstate, alba2015eigenstate, beugeling2015off, mondaini2017eigenstate, nation2018off, yoshizawa2018numerical, reimann2018dynamical, hamazaki2019random, khaymovich2019eigenstate, mierzejewski2020quantitative}, however, the rigorous proof of this hypothesis has remained elusive.

A popular approach to understanding the universal validity of the ETH is to invoke the typicality argument~\cite{neumann1929beweis}, which allows one to obtain a mathematically rigorous bound on the probability weight of the ETH-breaking Hamiltonians, thereby showing that an overwhelming majority of Hamiltonians satisfy the ETH~\cite{neumann1929beweis, goldstein2010long, goldstein2010approach, goldstein2010normal, reimann2015generalization}.
It is tempting to argue that most realistic Hamiltonians satisfy the ETH because most Hamiltonians do.
However, almost all Hamiltonians considered in Ref.~\cite{reimann2015generalization} involve nonlocal and many-body operators.
In fact, the typicality argument has recently been demonstrated to be inapplicable to a set of local Hamiltonians and local observables~\cite{hamazaki2018atypicality}. 

Another approach is to numerically test the ETH for physically realistic models involving local interactions between spins~\cite{steinigeweg2013eigenstate, kim2014testing, beugeling2014finite, steinigeweg2014pushing, alba2015eigenstate, beugeling2015off, mondaini2017eigenstate, nation2018off, hamazaki2019random, khaymovich2019eigenstate}, fermions~\cite{santos2010onset, khaymovich2019eigenstate, jansen2019eigenstate, mierzejewski2020quantitative} and bosons~\cite{santos2010onset, santos2010localization, ikeda2011eigenstate, ikeda2013finite, yoshizawa2018numerical, jansen2019eigenstate}.
This approach cannot clarify how generally the ETH applies to physical systems.
Indeed, recent studies have revealed exceptional systems for which the ETH breaks down: examples include systems with an extensive number of local conserved quantities~\cite{kinoshita2006quantum, rigol2007relaxation, rigol2009breakdown, iucci2009quantum, calabrese2011quantum, cassidy2011generalized, ilievski2015complete, hamazaki2016generalized, essler2016quench, vidmar2016generalized}, many-body localization (MBL)~\cite{basko2006metal, oganesyan2007localization, vznidarivc2008many, pal2010many, gring2012relaxation, nandkishore2015many, luitz2015many, choi2016exploring, smith2016many, imbrie2016many}, and quantum many-body scars~\cite{bernien2017probing, shiraishi2017systematic, turner2018weak, turner2018quantum, moudgalya2018exact, bull2019systematic, ho2019periodic, lin2019exact, schecter2019weak, shibata2020onsager}.

In this Letter, we present the first evidence that the ETH \textit{universally} holds true for \textit{locally} interacting quantum many-body systems.
We introduce random matrix ensembles constructed from local interactions and investigate their generic properties.
In particular, we evaluate the weight of the ETH-breaking Hamiltonians by numerically obtaining distributions of fluctuations of eigenstate expectation values~\cite{ref_local}.
We find that the ETH with exponentially small fluctuations is satisfied for an overwhelming majority of ensembles with local interactions. 
The obtained distribution shows that the fraction of exceptions is less suppressed for local ensembles than the conventional random matrix ensemble which involves nonlocal interactions and many-body interactions. 
Here, by many-body, we mean that the number of particles involved is comparable with the total number of particles.
If we allow less local interactions, the distribution rapidly approaches that predicted by the conventional random matrix theory.
We find that the ergodicity of our random matrix ensembles breaks down due to locality.

\paragraph{Local random matrix ensembles.}
We consider $N$ spins on a one-dimensional lattice with the periodic boundary condition and ensembles of Hamiltonians which contain only local interactions. 
We denote the local Hilbert space on each site as $\mathcal{H}_{\mathrm{loc}}$ and the total Hilbert space as $\mathcal{H}_{N} \coloneqq \mathcal{H}_{\mathrm{loc}}^{\otimes N}$. 
We choose an arbitrary orthonormal basis $\mathcal{B}_{\mathrm{loc}} = \Bqty*{ \ket*{\sigma} }$ of $\mathcal{H}_{\mathrm{loc}}$ and define the corresponding basis of $\mathcal{H}_{N}$ as $\mathcal{B}_{N} = \Bqty*{ \ket*{ \sigma_{1}\dots\sigma_{N} } \mid \forall j, \ \ket*{\sigma_{j}} \in \mathcal{B}_{\mathrm{loc}} }$.
The translation operator $\hat{T}_{N}$ acting on $\mathcal{H}_{N}$ satisfies
$
	\hat{T}_N \ket*{ \sigma_{1}\sigma_{2}\dots\sigma_{N} } \coloneqq \ket*{ \sigma_{2}\dots\sigma_{N}\sigma_{1} }
$
for all $\ket*{ \sigma_{1}\sigma_{2}\dots\sigma_{N} } \in \mathcal{B}_{N}$.

Let $\mathcal{L}(\mathcal{H})$ be the space of all Hermitian operators acting on a Hilbert space $\mathcal{H}$. 
For a given Hamiltonian $\hat{H}$, an energy shell $\mathcal{H}_{E,\delta E}$ centered at energy $E$ with width $2\delta E$ is defined as
$
  \mathcal{H}_{E,\delta E} \coloneqq \mathrm{span}\Bqty{ \ket*{E_{\alpha}} \mid \abs*{E_{\alpha} -E} \leq \delta E },
$
where $\ket*{E_{\alpha}}$ is an eigenstate of the Hamiltonian $\hat{H}$ with eigenenergy $E_{\alpha}$.
We randomly choose a local Hamiltonian $\hat{h}^{(l)}$ from the space $\mathcal{L}( \mathcal{H}_{\mathrm{loc}}^{\otimes l} )$ with respect to the Gaussian unitary ensemble.
We call an element of $\mathcal{L}( \mathcal{H}_{\mathrm{loc}}^{\otimes l} )$ an $l$-local operator and an integer $l \in \mathbb{N}$ the locality of an interaction. 
We define the range of the spectrum of an operator $\hat{O}$ as $\eta_{O} \coloneqq \max_{\alpha}a_{\alpha} -\min_{\alpha} a_{\alpha}$ where $a_{\alpha}$'s are eigenvalues of $\hat{O}$.
We consider Hamiltonians of the form
\begin{align}
  \hat{H}_{N} \coloneqq \sum_{j=0}^{N-1} \hat{T}_{N}^{j} \hat{h}^{(l)}_{j}	\hat{T}_{N}^{-j},   \label{Ensemble:Hamiltonian}
\end{align}
which respect the locality of interactions, and introduce the following three types of ensembles with different restrictions~\cite{case2_local}:
{ \setlength{\leftmargini}{50pt} 
\begin{enumerate}
    \renewcommand{\labelenumi}{Case \arabic{enumi}:}
    \item $h^{(l)}_{j} = h^{(l)}$ for all $j$.
    \item $h^{(l)}_{j}$ is normalized so that $\eta_{h^{(l)}_{j}} = \eta$ for all $j$.
    \item No restrictions.
\end{enumerate}
}
\noindent
The number of parameters needed to characterize a single Hamiltonian increases from Case~1 to Case~3~\cite{Levy_lemma}.
Observables are randomly chosen with respect to Case~1.
That is, we randomly choose an $l$-local observable $\hat{o}^{(l)} \in \mathcal{L}(\mathcal{H}_{\mathrm{loc}}^{\otimes l})$ from the Gaussian unitary ensemble and construct an extensive observable $\hat{O}_{N} \in \mathcal{L}(\mathcal{H}_{N})$ as in 
Eq.~\eqref{Ensemble:Hamiltonian} with $\hat{o}^{(l)}_{j} = \hat{o}^{(l)}$ for all $j$.

\paragraph{Measure of the strong ETH.}
We focus on the strong ETH, which asserts that all eigenstates be thermal.
While several definitions for a measure~\cite{biroli2010effect, santos2010localization, steinigeweg2013eigenstate, beugeling2014finite, de2015necessity, reimann2015generalization, hamazaki2018atypicality} of the ETH has been proposed, we consider a measure that is also applicable to generic local systems.
We require that the measure be 
    (i) invariant under the linear transformation$\colon \hat{H} \mapsto a\hat{H}+b, \ \hat{O} \mapsto a' \hat{O} +b'$,
    (ii) dimensionless,
    (iii) thermodynamically intensive,
 and (iv) applicable to eigenstate expectation values after the subtraction of weak energy dependences.
Here, (i) is needed because the measure of the strong ETH should be invariant under a change of physical units and under translation of the origin of physical quantities;
(ii) is needed because we compare quantities with different physical dimensions;
and (iii) is needed because we admit subextensive fluctuations from a macroscopic point of view.
Lastly, (iv) is important because the energy dependence generically appears in the presence of locality of interactions.
Such a dependence invalidates the typicality argument based on a unitary Haar measure unless the energy width is exponentially small~\cite{hamazaki2018atypicality}.
Since this energy dependence of a macroscopic observable can be observed, it should not be considered to be a part of fluctuations of eigenstate expectation values.
To be concrete, consider 
a measure of the strong ETH as
$
	\tilde{\Delta}_{\infty} \coloneqq { \displaystyle\max_{\alpha\colon \ket*{ E_{\alpha} } \in \mathcal{H}_{E,\delta E} } \abs*{ O_{\alpha\alpha} -\expval*{ \hat{O} }^{\mathrm{mc}}_{\delta E}(E)} }/{ \eta_{O} },
$
where $\expval*{\cdots}_{\delta E}^\mathrm{mc}(E)$ is the microcanonical average within $\mathcal{H}_{E,\delta E}$.
This is essentially the same quantity as that used in Ref.~\cite{reimann2015generalization}.
The scaling behavior of this measure depends on the scaling of the energy width $\delta E$, which is inappropriate as the measure of the strong ETH. 
Such an energy dependence is removed if we consider eigenstate-dependent microcanonical energy shell and introduce the following measure:
\begin{align}
	\Delta_{\infty} &\coloneqq \frac{ \max_{\alpha} \abs*{ O_{\alpha\alpha} -\expval*{ \hat{O} }^{\mathrm{mc}}_{\delta E}(E_{\alpha})} }{ \eta_{O} }, \label{Eq:Measure}
\end{align}
where the maximum is taken over the middle 10\% of the energy spectrum to avoid finite-size effects at both edges of the spectrum where the density of states is small. 
The strong ETH implies that $\Delta_{\infty}\rightarrow 0$ in the thermodynamic limit.

To analytically test the strong ETH is formidable since we do not know exact statistics of eigenstates of a general Hamiltonian $\hat{H}_{N}$.
Therefore, we employ the exact diagonalization method to investigate the universality of the ETH.
For Case~1, where translation invariance exists, we restrict ourselves to the zero-momentum sector.
An analytical method based on a uniform-random-vector method over the Haar measure~\cite{reimann2015generalization} can no longer be applied to our Hamiltonians because of their local structures~\cite{hamazaki2018atypicality}.

\paragraph{Strong ETH for almost all local random matrices.}
We numerically obtain the distributions of $\Delta_{\infty}$ for several system sizes $N$ and locality $l$. 
We first demonstrate that the ETH holds true for almost all local random matrices in terms of three cases of ensembles defined above on the basis of Markov's inequality:
\begin{align}
  \mathrm{Prob}^{(l)}_{N}\bqty*{ \Delta_{\infty} \geq \epsilon } \leq \frac{ \mathbb{E}_{N}^{(l)}[\Delta_{\infty}] }{ \epsilon }. \label{eq:Markov}
\end{align}
Here, $\mathrm{Prob}$ and $\mathbb{E}$ denote the probability and the expectation with respect to random realizations of $\hat{H}$ and $\hat{O}$.
The vanishing of $\mathbb{E}_{N}^{(l)}[\Delta_{\infty}]$ in the thermodynamic limit is a sufficient condition for the strong ETH with an arbitrary constant $\epsilon > 0$ for almost all sets of local Hamiltonians and observables.

\begin{figure}[tb]
    \centering
    \includegraphics[width=\linewidth]{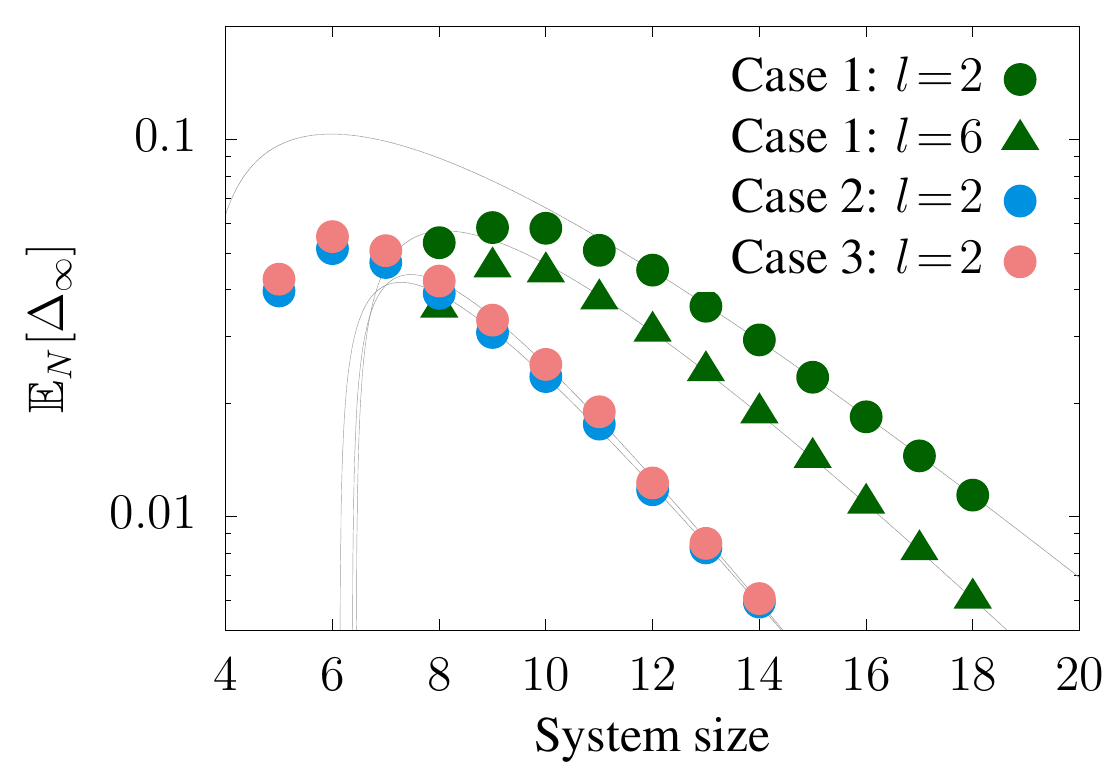}
	\caption{\label{Figure:Mean} Mean value of the measure in Eq.~\protect\eqref{Eq:Measure} for the strong ETH $\mathbb{E}_{N}\bqty*{\Delta_{\infty}}$ for various ensembles.
	The solid curve is the fitting function in Eq.~\eqref{eq:MeanTI} for Case~1 or Eq.~\eqref{eq:MeanNonTI} for Cases~2 and 3 (see Supplemental Material~\protect\cite{supplement_local}). 
	The values of the fitting parameters $(N_{m}, N_{0}, m_{0})$ are $(3.20, 2.90, 0.21)$ for Case~1 with $l=2$, $(2.71, 5.26, 0.33)$ for Case~1 with $l=6$, $(2.29, 6.13, 0.94)$ for Case~2 with $l=2$, and $(2.20, 6.37, 1.25)$ for Case~3 with $l=2$.
	The values $N_{m}$ for Cases~2 and 3 are smaller than the expected value $2/\log 2$ due to a finite-size effect.
	The number of samples lies between 7980 and 947770 for all data points.
	}
\end{figure}

We compare our numerical results with the prediction of
conventional random matrix theory, whose asymptotic $N$-dependence is obtained as (see discussions after Eq.~\eqref{eq:CDF})
\begin{align}
  \mathbb{E}_{N}[\Delta_{\infty}] &= m_{0} N e^{ -N/N_{m} } \sqrt{ 1-\frac{ N_{m} }{2} \frac{ \log N }{N} -\frac{ N_{0} }{N} } \label{eq:MeanTI}
\end{align}
for Case~1 and
\begin{align}
  \mathbb{E}_{N}[\Delta_{\infty}] &= m_{0} N^{1/2} e^{ -N/N_{m} } \sqrt{ 1 -\frac{ N_{0} }{N} } \label{eq:MeanNonTI}
\end{align}
for Cases~2 and 3, where $m_{0}, N_{0}$ and $N_{m}$ are some constants.
As shown in Fig.~\ref{Figure:Mean}, these formulas fit well to our numerical data for all the ensembles irrespective of locality $l$.
While Eqs.~\eqref{eq:MeanTI} and \eqref{eq:MeanNonTI} are expected to apply to a less local case (i.e., $l$ is large) with not too small system sizes, where $\mathrm{Prob}^{(l)}_{N}\bqty*{ \Delta_{\infty} \geq \epsilon }$ itself is close to that for the conventional random matrix theory, they fit well to the cases with strong locality ($l=2$), where  $\mathrm{Prob}^{(l)}_{N}\bqty*{ \Delta_{\infty} \geq \epsilon }$ is distinct from that of conventional random matrix theory.

The exponential decay of $\mathbb{E}_{N}^{(l)}[\Delta_{\infty}]$ allows one to make both $\epsilon$ and the right-hand side of Eq.~\eqref{eq:Markov} exponentially small by taking $\epsilon_{N} \propto \exp( -N/N_{1} )$ with $N_{1} > N_{m}$. 
This means that the strong ETH with exponentially small fluctuations ($\sim \exp(-N/N_1)$) holds for an overwhelming majority of the ensemble, where the fraction of exceptional cases is exponentially small~\cite{edge_local}.
We also note that $N_{m}$ is close to $2/\log 2$, since the standard deviation of $O_{\alpha\alpha}$ decays as $1/\sqrt{d_{\mathrm{sh}}}$~\cite{srednicki1999approach} and  $d_{\mathrm{sh}} \propto \dim\mathcal{H}_{N}=2^N$ where $d_{\mathrm{sh}} \coloneqq \dim\mathcal{H}_{E,\delta E}$ irrespective of $l$ unless $\delta E$ decreases exponentially with $N$~\cite{supplement_local}.

Notably, the ETH universally holds even for Cases 2 and 3.
Our results show that MBL rarely occurs for these types of spatial disorder.
This is similar to the many-body chaos found in random unitary circuits~\cite{chan2018spectral, chan2018solution}; however our results further suggest that the randomness does not prevent thermalization even with energy conservation due to continuous time evolution.
We also test the argument that localization may occur when the magnitude of the sum of off-diagonal elements of a Hamiltonian exceeds the magnitude of a diagonal one. Assuming independent Gaussian elements, we estimate the probability that a sample may show MBL to be $\sim \exp(-\order{N d_{\mathrm{loc}}^{\, l} })$ (see Supplemental Material~\cite{supplement_local}).
However, since off-diagonal elements of our Hamiltonians are highly correlated due to the spatial locality, the relevance of the above estimate remains unclear.

\begin{figure}
    \centering
    \includegraphics[width=\linewidth]{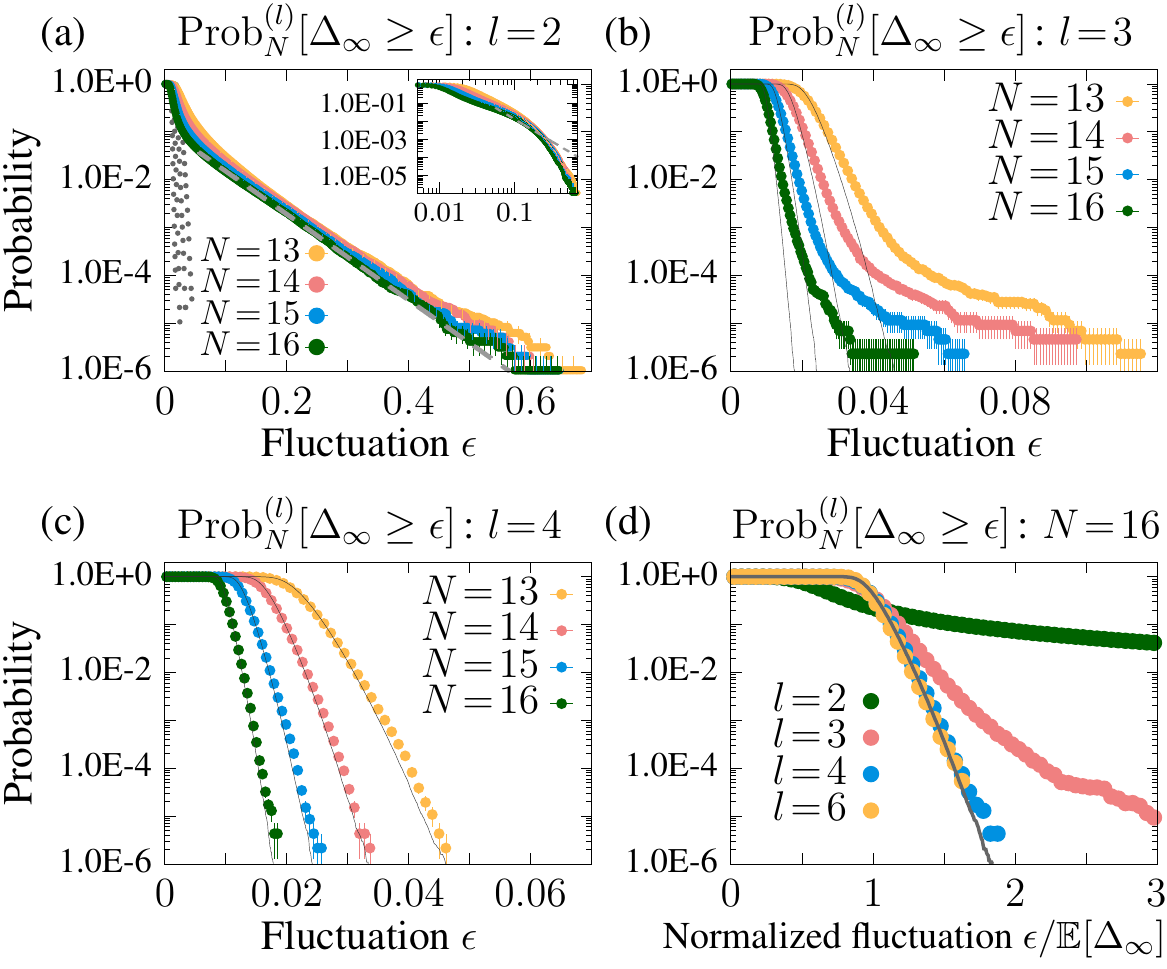}
    \caption{\label{Figure:TI_Distribution}
    (a) $\mathrm{Prob}_{N}^{(l)}\bqty*{ \Delta_{\infty} \geq \epsilon }$ with $l=2$ as a function of $\epsilon$ for various system sizes (colored dots).
	The distributions for $l=6$ (gray dots) with $N=13\!\sim\!16$ are shown for comparison.
	The gray dashed lines show exponential functions of the form $C\exp(-\epsilon/\epsilon_{0})$.
    The dashed line in the inset (log-log plot) shows a polynomial function of the form $(\epsilon_{1}/\epsilon)^{a}$.
	The tail fittings are performed in the region $\epsilon > 3\mathbb{E}_{N}\bqty*{\Delta_{\infty}}$.
	The number of samples is 947770.
	(b) and (c) show $\mathrm{Prob}_{N}^{(l)}\bqty*{ \Delta_{\infty} \geq \epsilon }$ with $l=3$ and $4$, respectively.
	(d) Distribution of $\Delta_{\infty}$ normalized with $\mathbb{E}_{N}^{(l)}\bqty*{\Delta_{\infty}}$ for Case~1 with $l=2,3,4,6$ and $N=16$.
	The gray line is a maximum value distribution predicted from the conventional random matrix theory, which is rescaled so that its mean becomes unity.
    }
\end{figure}

\paragraph{Distribution of the maximum fluctuation.}
Since Markov's inequality gives only a loose upper bound, we directly obtain distributions of $\Delta_{\infty}$ for several values of $N$.
Below, we focus on Case~1 (see Supplemental Material for Cases 2 and 3~\cite{supplement_local}).
The results are shown in Fig.~\ref{Figure:TI_Distribution}, where the distribution with $l=2$ (Fig.~\ref{Figure:TI_Distribution}(a)) is distinct from the prediction of the conventional random matrix theory involving nonlocal operators. 
We find that its tail decays single-exponentially or slightly slower than a single-exponential $\exp(-\epsilon/\epsilon_{1})$ unlike the conventional random matrix theory which predicts a much faster decay of the tail as $\exp( -\order*{\epsilon^2} )$.
This suggests that locality favors Hamiltonians with relatively large $\Delta_{\infty}$.
This can be attributed to the closeness to those Hamiltonians that are integrable or host scars.
The distribution of $\Delta_{\infty}$ for Case~1 with $l=3$ shows a crossover from a rapid decay in the region $\mathrm{Prob}_{N}[\Delta_{\infty} \geq \epsilon] \gtrsim P_c\sim 1.0 \mathrm{E}{-4}$ followed by a slower decay in the region $\mathrm{Prob}_{N}[\Delta_{\infty} \geq \epsilon] \lesssim P_c\sim 1.0\mathrm{E}{-4}$ (Fig.~\ref{Figure:TI_Distribution}(b)).
This behavior is similar to the case with $l=2$ but $P_c$ is much smaller ($P_c\sim 1.0 \mathrm{E}{-2}$ for $l=2$).

As the locality $l$ increases, the distributions of $\Delta_{\infty}$ rapidly approach the prediction of the conventional random matrix theory, where the fluctuations of eigenstates distribute according to the Gaussian distribution with zero mean and the identical variance $s_{N}^2$ for each sample. 
Indeed, as shown in Figs.~\ref{Figure:TI_Distribution}(c) and (d), even for $l$ as small as $4$, $\mathrm{Prob}^{(l)}_{N}[\Delta_\infty\geq\epsilon]$ is well fitted by the cumulative function of the maximum absolute value of $d_{\mathrm{sh}}$-independent and identically distributed Gaussian variables,
\begin{align}
    \mathrm{Prob}_{N}^{(\mathrm{RMT})}\bqty*{ \Delta_{\infty} \geq \epsilon } 
    &= 1- \mathrm{CDF}(\epsilon) \nonumber \\
    &= 1-\bqty{ \mathrm{erf}(\epsilon/\sqrt{2s_{N}^2}) }^{ d_{\mathrm{sh}} }, \label{eq:CDF}
\end{align}
where $\mathrm{erf}(x)$ is the error function and $d_{\mathrm{sh}} \coloneqq \dim\mathcal{H}_{E,\delta E}$.

The extreme value theory~\cite{de2007extreme} allows us to obtain the asymptotic form of the cumulative distribution function:
if we set $b_{N} \!\sim\! s_{N} \sqrt{2\log d_{\mathrm{sh}}}$ and $a_{N} \!=\! s_{N}^2/b_{N}$, the right-hand side in Eq.~\eqref{eq:CDF} converges to the Gumbel distribution 
$
\mathrm{Prob}^{( \mathrm{RMT} )}_{N}\bqty*{ \Delta_{\infty} \!\geq\!\epsilon }\!\sim\! 1\!-\!\exp[-e^{-\frac{\pi}{\sqrt{6}} y -\gamma}]
$
for large $d_\mathrm{sh}$, where $y \coloneqq (\epsilon -b_{N}) / a_{N}$ is a rescaled random variable, and $\gamma \simeq 0.577$ is the Euler-Mascheroni constant~\cite{supplement_local}.
This fact implies that $\mathbb{E}_{N}[\Delta_\infty] \simeq b_N \!\sim\! s_N\sqrt{2\log d_\mathrm{sh}}$ and $\mathbb{S}_{N}[\Delta_\infty] \simeq a_N\!\sim\! s_N/\sqrt{2\log d_\mathrm{sh}}$.
This distribution is applicable in the range $\epsilon = \mathbb{E}_{N}[\Delta_{\infty}] +c\, \mathbb{S}_{N}[\Delta_{\infty}]$ where $c$ is a constant of $\order{1}$ with respect to $N$.

These formulas lead to the asymptotic $N$-dependence of $\mathbb{E}_{N}[\Delta_{\infty}]$ in the conventional random matrix regime.
Since $d_{\mathrm{sh}}\propto \dim\mathcal{H}_{N}$ and $s_{N} \propto (d_{\mathrm{sh}})^{-1/2}$ for sufficiently large $N$,
we obtain the asymptotic formulas in Eqs.~\eqref{eq:MeanTI} and \eqref{eq:MeanNonTI} by inserting $\dim\mathcal{H}_{N} = d_{\mathrm{loc}}^{\, N}/N$ for Case~1 and $\dim\mathcal{H}_{N} = d_{\mathrm{loc}}^{\, N}$ for Cases 2 and 3 in $\mathbb{E}_{N}[\Delta_\infty] \!\sim\! s_N\sqrt{2\log d_\mathrm{sh}}$.

\paragraph{Ergodicity breaking for local random matrices.}
Let us now discuss the entire structure of the expectation values over eigenstates for random realizations of sets of local Hamiltonians and observables.

Srednicki conjectured~\cite{srednicki1999approach} that
the fluctuations of expectation values can be expressed as $\delta (O_{N})_{\alpha\alpha}\sim e^{ -\frac{ S_{N}(E) }{2} } f_{O}(E) \tilde{R}_{\alpha\alpha}$.
Here, $S_{N}(E)$ is the thermodynamic entropy of the system which only depends on the Hamiltonian $\hat{H}_{N}$, $f_{O}(E)$ is a smooth function of energy $E$ that depends on $\hat{H}_{N}$ and $\hat{O}_{N}$,
and $\tilde{R}_{\alpha\alpha}$ distributes according to the normal Gaussian distribution.

\begin{figure}[tb]
    \centering
    \includegraphics[width=0.48\linewidth]{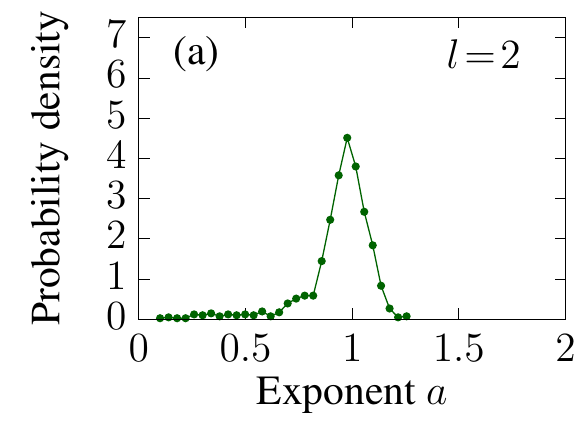}
    \includegraphics[width=0.48\linewidth]{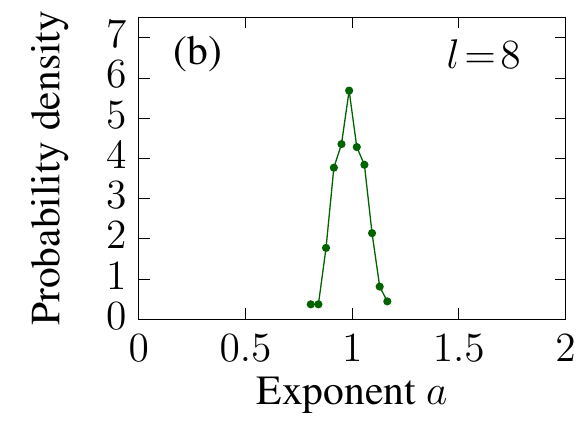}
	\caption{\label{Figure:SredniciAnsatz}
	Distribution of the value of $a$ in the fitting of the standard deviation $\mathcal{S}_{\gamma}^{(E,\delta E)}[\delta(O_{N})_{\gamma\gamma}] \propto (\sqrt{ d_{\mathrm{sh}} })^{-a}$ in the shell $\mathcal{H}_{E,\delta E}$ for the Case~1 ensemble with (a)~$l=2$ and (b)~$l=8$, where $E$ is chosen to be the center of the spectrum and $\delta E$ is 5\% of the spectral range.
    The number of samples is (a)~1022 and (b)~379.
	}
\end{figure}

We test the above conjecture for our three ensembles for $l=2$ as a local case and $l=8$ as a nonlocal case.
We find that Srednicki's conjecture typically holds irrespective of the locality, that is, the standard deviation of $\delta(O_{N})_{\alpha\alpha}$ inside an energy shell $\mathcal{H}_{E,\delta E}$ typically scales as $\propto (\sqrt{d_{\mathrm{sh}}})^{-a}\sim e^{-a\frac{S_N(E)}{2}}$ with $a\simeq 1$ (Fig.~\ref{Figure:SredniciAnsatz}), and $\tilde{R}_{\alpha\alpha}$ distributes according to the normal Gaussian~\cite{supplement_local}.

We also find that sample-to-sample fluctuations become large for local random matrices.
The ergodicity of a random matrix ensemble~\cite{guhr1998random}, which means that the spectral average equals the ensemble average, does not apply to real situations with locality.
We observe its signature in Fig.~\ref{Figure:SredniciAnsatz}(a), where samples with small $a$ exist for $l=2$, while the distribution concentrates around $a=1$ for a less local case with $l=8$.
We find that atypical samples with small $a$ have multi-fractal eigenstates even in the middle of the spectrum~\cite{supplement_local}.

\begin{figure}[tb]
    \centering
    \includegraphics[width=0.48\linewidth]{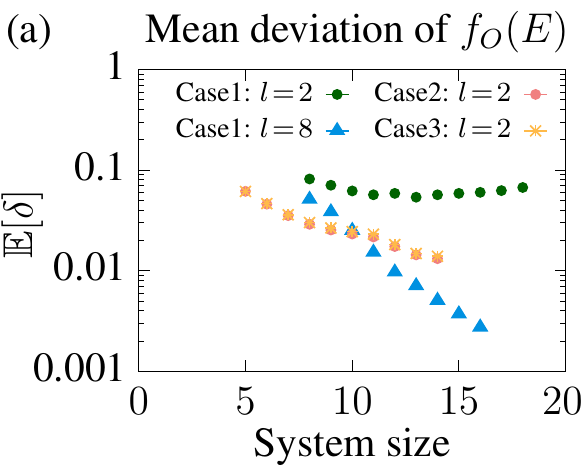}
    \includegraphics[width=0.48\linewidth]{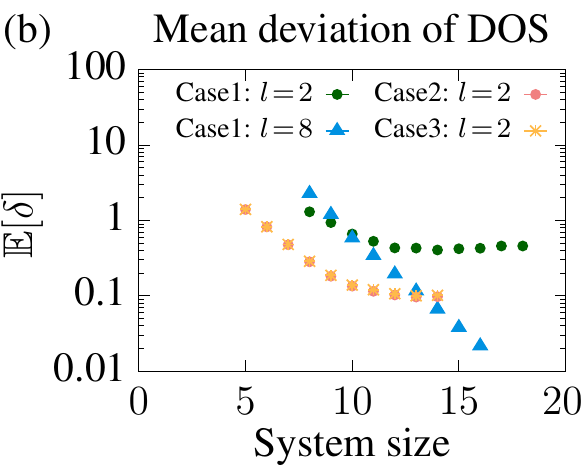}
	\caption{\label{Figure:Structures} 
	Mean $\mathbb{E}\bqty*{\delta}$ of the $L$2-norm from the ensemble averages for (a) $f_{O}(E)$ in Srednicki's conjecture and (b) the normalized density of states $\rho(E)$. 
	The number of samples is 48800 for Case~1 with $l=2$, 43293 for Case~1 with $l=8$, and 10000 for Case~2 and Case~3.
	}
\end{figure}

Figure~\ref{Figure:Structures} shows the mean of the $L$2-norm $\delta$ defined by
\begin{align}
    \delta \coloneqq \bqty{ \frac{1}{ N_{\mathrm{bin}} } \sum_{\mathrm{bin}=1}^{ N_{\mathrm{bin}} } \qty( \expval*{X(E_\alpha)}_{\mathrm{bin}} - \mathbb{E}\bqty*{ \expval*{X(E_\alpha)}_{\mathrm{bin}} } )^2 }^{1/2},
\end{align}
where $\expval*{\cdots}_{\mathrm{bin}}$ denotes the average inside each bin and $\mathbb{E}\bqty*{\cdots}$ denotes the ensemble average.
Figures~\ref{Figure:Structures}(a) and (b) show the $N$-dependence of $X(E)=f_{{O}}(E)$ and the normalized density of states $X(E)=\rho(E)$ (i.e., the density of states divided by $\dim\mathcal{H}_{N}$), respectively, for $N_{\mathrm{bin}} = 100$.
The mean of the $L$2-norm for $f_{O}(E)$ decreases with $N$ for the nonlocal ensemble (Case~1, $l=8$), which is a manifestation of the ergodicity in the conventional random matrix ensemble~\cite{guhr1998random}.
On the other hand, it converges to some finite value for local ensembles with $l=2$ for Cases 1, 2, and 3.
The breakdown of ergodicity for local ensembles can also be seen in the density of states. 

\paragraph{Conclusion.}
We find that the locality of interactions on an ensemble of Hamiltonians makes the distributions of local observables significantly different from those of the conventional random matrix theory.
However, the strong ETH with exponentially small fluctuations holds true for an overwhelming majority of the ensemble, where the fraction of exceptions is exponentially small. 
We also find that the ergodicity of random matrix ensembles breaks down due to locality.
We expect that the universality of the ETH still holds true for higher dimensions, since integrability such as the Bethe-ansatz-solvable one is unique to 1D and MBL seems unstable in higher dimensions~\cite{de2014scenario}.

While the universality of the ETH is confirmed in all three ensembles studied here, it is of fundamental interest to investigate whether imposing additional conserved quantities can prevent the universality.
It is also of interest to examine whether the ETH-MBL transition occurs if we implement more structured randomness than the Case~2 and Case~3 ensembles, such as ensembles where the strengths of 1-site and 2-site disorder are different.
Our ensembles can provide a relaxation timescale of generic interacting Hamiltonians with locality, which is not taken into account in related works~\cite{reimann2016typical, balz2017typical, reimann2019transportless}. 

\begin{acknowledgments}
We thank Zongping Gong for pointing out the importance of Markov's inequality and Takashi Mori for pointing out a subtlety concerning the measure of the strong ETH.
This work was supported by KAKENHI Grant Numbers JP19J00525 and JP18H01145, and a Grant-in-Aid for Scientific Research on Innovative Areas (KAKENHI Grant Number JP15H05855) from Japan Society for the Promotion of Science (JSPS). 
S.~S. was supported by Forefront Physics and Mathematics Program to Drive Transformation (FoPM), a World-leading Innovative Graduate Study (WINGS) Program, the University of Tokyo.
R.~H. was supported by JSPS through Program for Leading Graduate Schools (ALPS) and JSPS fellowship (KAKENHI Grant Number JP17J03189).
\end{acknowledgments}

\bibliography{local}

\begin{thebibliography}{87}
\bibitem [{ref()}]{ref_local}%
  \BibitemOpen
  \href@noop {} {}\bibinfo {note} {{We note that Ref.~\cite{brandino2012quench}
  addresses the ETH for random matrices with sparsity, which is a property of a
  Hamiltonian with only local interactions, for some concrete examples;
  however, it does not discuss statistics over different samples and hence the
  universality of the ETH.}}\BibitemShut {Stop}%
\bibitem [{Lev()}]{Levy_lemma}%
  \BibitemOpen
  \href@noop {} {}\bibinfo {note} {{The cardinality of the ensemble is
  independent of the system size $N$ for Case~1, while it increases
  exponentially with $N$ for Case~2 and Case~3. Therefore, we cannot apply the
  results of measure concentration such as Levy's lemma especially to
  Case~1.}}\BibitemShut {Stop}%
\bibitem [{edg()}]{edge_local}%
  \BibitemOpen
  \href@noop {} {}\bibinfo {note} {{These results do not contradict those of
  Ref.~\cite{brandino2012quench}, which states that the strong ETH fails near
  the edge of the spectra in the presence of the locality, because we focus on
  the middle of the spectra.}}\BibitemShut {Stop}%
\end{thebibliography}

\end{document}


\title{
    Supplemental Material: \protect\\
    Test of Eigenstate Thermalization Hypothesis Based on Local Random Matrix Theory
}
\author{Shoki Sugimoto}
    \affiliation{Department of Physics, University of Tokyo, 7-3-1 Hongo, Bunkyo-ku, Tokyo 113-0033, Japan}
\author{Ryusuke Hamazaki}
    \affiliation{Department of Physics, University of Tokyo, 7-3-1 Hongo, Bunkyo-ku, Tokyo 113-0033, Japan}
    \affiliation{Nonequilibrium Quantum Statistical Mechanics RIKEN Hakubi Research Team, RIKEN Cluster for Pioneering Research (CPR), RIKEN iTHEMS, Wako, Saitama 351-0198, Japan}
\author{Masahito Ueda}
    \affiliation{Department of Physics, University of Tokyo, 7-3-1 Hongo, Bunkyo-ku, Tokyo 113-0033, Japan}
    \affiliation{RIKEN Center for Emergent Matter Science (CEMS), Wako 351-0198, Japan}
\date{\today}
\maketitle

\section{Derivation of the Gumbel distribution}
\label{Appendix:Derivation of the Gumbel distribution}

Let $X_{1},X_{2},\dots, X_{n}$ be $n$ independent and identically distributed (i.i.d) Gaussian variables with zero mean and variance $s_{n}^2$. 
The cumulative distribution function of the maximum of the absolute value of $X_{i}$'s is $\mathrm{CDF}_{n}(x) = \bqty*{ \mathrm{erf}(x/\sqrt{2s_{n}^2}) }^{n}$. Let us introduce a random variable $y$ by $x/s_{n} = a_{n}y +b_{n}$ for some sequences $\Bqty*{a_{n}}$ and $\Bqty*{b_{n}}$. 
It is known from the extreme value theory that if we choose the sequences $\Bqty*{a_{n}}$ and $\Bqty*{b_{n}}$ appropriately, the distribution of $y$ converges to some universal distribution~\cite{fisher1928limiting}. 
We determine such sequences below.
Since the error function has an asymptotic form $\mathrm{erf}(x) \sim 1 -\frac{1}{\sqrt{\pi} x} e^{-x^2}$ for $\abs*{x}\gg 1$, we have
\begin{align}
    \mathrm{CDF}_{n}(x) = \qty( 1 - \sqrt{ \frac{2}{\pi} } \frac{1}{a_{n}y +b_{n}} e^{-\frac{ \qty(a_{n}y +b_{n})^2 }{2} } )^n
\end{align}
for $\abs*{x} \gg s_{n}$.
If we choose $\Bqty*{b_{n}}$ to satisfy $e^{-b_{n}^2/2}/b_{n} = A_{1}/n$ and $a_{n} = A_{2}/b_{n}$ for some constants $A_{1}, A_{2} > 0$, we have $b_{n} \simeq \sqrt{2 \log n}$ for large $n$. 
For our purpose, it is sufficient to calculate $b_{n}^2$ up to terms of $O(1)$.
If we define $\delta$ by $\delta \coloneqq b_{n} -\sqrt{2 \log n}$, where $\delta / \sqrt{2 \log n} \to 0$ for $n\to\infty$, we need to calculate $\delta$ up to terms of $O(1) / \sqrt{2 \log n}$.
For this purpose, we first expand $\log b_n$ in terms of $\delta/\sqrt{2\log n}$, obtaining
\begin{align}
    \log \frac{A_{1}}{n} &= -\frac{1}{2} b_{n}^2 -\log b_{n} \nonumber \\
    &= -\log n -\delta \sqrt{2\log n} -\frac{1}{2} \delta^2 -\frac{1}{2} \log( 2\log n ) -\log( 1 +\frac{ \delta }{ \sqrt{ 2\log n } } ) \nonumber \\
    &= \log\frac{1}{n} - (2\log n) \bqty{ \frac{ \delta }{ \sqrt{2\log n} } +\frac{1}{2}\frac{ \log(2\log n) }{ 2\log n } +\frac{1}{2} \qty(\frac{ \delta }{ \sqrt{2\log n} } )^2 +\order{ \frac{ \delta }{ (2\log n)^{3/2} } } }. \label{Eq:expandB}
\end{align}
This equation shows that the leading term of $\delta$ is $-\log(2\log n) / 2\sqrt{2\log n}$.
Thus, the last two terms in Eq.~(S-\ref{Eq:expandB}) can be ignored. 
We therefore obtain
\begin{align}
    \delta = -\frac{1}{2} \frac{ \log(2\log n) }{ \sqrt{2\log n} } -\frac{ \log A_{1} }{ \sqrt{2\log n} } +o\qty( \frac{1}{\sqrt{2\log n}} ).
\end{align}
Finally, we have
\begin{align}
    b_{n}^2 = 2\log n -\log( 2\log n ) -2\log A_{1}+o(1).
\end{align}
Since $a_{n} = A_{2}/b_{n}$, $a_{n} \to 0$ for large $n$ and hence
\begin{align}
    \mathrm{CDF}_{n}(x) &= \qty( 1 - \sqrt{ \frac{2}{\pi} } \frac{1}{a_{n}y +b_{n}} e^{-\frac{ \qty(a_{n}y +b_{n})^2 }{2} } )^n = \qty( 1 -\sqrt{ \frac{2}{\pi} } \frac{b_{n}}{ a_{n}y +b_{n} } e^{ -\frac{ a_{n}^2y^2 }{2} -a_{n}b_{n} y } \frac{ e^{ -\frac{b_{n}^2}{2} } }{b_{n}} )^n \nonumber \\
    &= \qty( 1 -\sqrt{ \frac{2}{\pi} } \frac{b_{n}}{ a_{n}y +b_{n} } \frac{A_{1}}{n} e^{ -\frac{ a_{n}^2y^2 }{2} -A_{2} y } )^n \nonumber \\
    &\xrightarrow{ n\to\infty } \exp( -A_{1}\sqrt{ \frac{2}{\pi} } e^{-A_{2}y} ) \quad\text{(Gumbel distribution).}
    \label{Eq:Gumbel}
\end{align}
In the last line, we assume that $y$ is independent of $n$, and thus the convergence is pointwise in terms of $y$. 
This condition means that $x/s_n \simeq b_n = \order{ \sqrt{\log n} }$ from the relation $x/s_n = a_n y + b_n$ and $a_n \to 0$ for $n\to \infty$.
This distribution is called a Gumbel distribution. 
Therefore, the distribution of $y$ becomes independent of $n$ for sufficiently large $n$. 

We here introduce quantities $m\coloneqq A_{2} \mathbb{E}[y] -\log (A_{1}\sqrt{2/\pi})$ and $s=A_{2}\mathbb{S}[y]$, which are shown to be independent of $A_{1}$ and $A_{2}$ from Eq.~\eqref{Eq:Gumbel}. 
Indeed, $m$ is the Euler-Mascheroni constant ($m \simeq 0.577$) and $s=\pi/\sqrt{6}$. 

It follows from the above arguments that
\begin{align}
    \mathbb{E}\bqty*{x/s_{n}} &= a_{n}\mathbb{E}[y] +b_{n} = \frac{ m+\log(A_{1} \sqrt{2/\pi}) }{ b_{n} } + b_{n} \sim \sqrt{ 2\log{n} }\qc \label{Eq:Mean} \\
    \mathbb{S}\bqty*{x/s_{n}} &= a_{n} \mathbb{S}[y] = \frac{s}{ b_{n} } \sim s/\sqrt{ 2\log{n} }. \label{Eq:Stddev}
\end{align}
The first relation together with $s_{N} \propto (d_{\mathrm{sh}})^{-1/2}$ and $n = d_{\mathrm{sh}} \propto d_{\mathrm{loc}}^{\, N}/N$ or $n = d_{\mathrm{sh}} \propto d_{\mathrm{loc}}^{\, N}$ yields Eqs.~(4) and (5) in the main text, respectively.

By eliminating $A_{2}y -\log(A_{1}\sqrt{2/\pi})$ using the relations $x/s_{n} = a_{n} y +b_{n}$, $a_{n} b_{n} = A_{2}$, and Eq.~\eqref{Eq:Stddev}, we obtain the cumulative distribution function of  $x$ near its mean value $\mathbb{E}[x]$ as
\begin{align}
    \mathrm{CDF}_{n}(x) &\xrightarrow{ n\to\infty } \exp( -e^{- \frac{\pi}{\sqrt{6}} \frac{ x -\mathbb{E}[x]} { \mathbb{S}[x] } -\gamma } ),
\end{align}
which is presented below Eq.~(6) in the main text.

To compare the above expression with the numerically obtained distribution of $\Delta_{\infty}$ with $l=6$, we set $n = d_{\mathrm{sh}} (\eqqcolon d_{0}e^{ N/N_{\mathrm{sh}} })$, which increases exponentially in $N$.
This means that, if we set $\epsilon = c\mathbb{E}\bqty*{x} \ (c>1)$ and rewrite $s_{d_{\mathrm{sh}}}$ as $s_{N}$, we have
\begin{align}
    \mathrm{Prob}_{N}\bqty\Big{ x \geq c\mathbb{E}\bqty*{x} } &= 1-\mathrm{CDF}_{ d_{\mathrm{sh}} }( c\mathbb{E}\bqty*{x} ) \nonumber \\
    &= 1-\qty( 1 -\sqrt{ \frac{2}{\pi} } \frac{1}{ c\mathbb{E}\bqty*{x/s_{N}} } e^{ -\frac{ (c \mathbb{E}\bqty*{x/s_{N}} )^2 }{2} } )^{ d_{\mathrm{sh}} } \nonumber \\
    &\simeq \sqrt{ \frac{2}{\pi} } \frac{ d_{\mathrm{sh}} }{ c\mathbb{E}\bqty*{x/s_{N}} } e^{ -\frac{ (c\mathbb{E}\bqty*{x/s_{N}})^2 }{2} }, \label{Eq:AppendixA}
\end{align}
where $ c\mathbb{E}\bqty*{x/s_{N}} \gg 1$.
In the third line, we use an expansion $(1-z)^{d} \simeq 1-zd$, which requires the smallness of the quantity $zd$.
Therefore, the last term of Eq.~\eqref{Eq:AppendixA} must be small. 
We note that we cannot apply Eq.~\eqref{Eq:Gumbel} since the difference $x/s_{n} -b_{n} \sim (c-1) \mathbb{E}[x]/s_{n}$ and hence $y$ becomes large in this case.
We simplify this expression to find the $N$-dependence.
From Eq.~\eqref{Eq:Mean}, we have
\begin{align}
    \mathbb{E}\bqty*{x/s_{N}}^2 &= b_{n}^2 +2m +2\log\qty(A_{1} \sqrt{ \frac{2}{\pi} } ) +o(1) \nonumber \\
    &= 2\log n -\log(2\log n) +2\gamma +2\log \sqrt{ \frac{2}{\pi} } +o(1).
\end{align}
We use this result to obtain
\begin{align}
    \mathrm{Prob}_{N}\bqty\Big{ x \geq c\mathbb{E}\bqty*{x} } 
    &\simeq \sqrt{ \frac{2}{\pi} } \frac{ d_{\mathrm{sh}} }{ c\mathbb{E}\bqty*{x/s_{N}} } e^{ -\frac{ (c\mathbb{E}\bqty*{x/s_{N}})^2 }{2} } \nonumber \\
    &= \sqrt{ \frac{2}{\pi} } \frac{ d_{\mathrm{sh}} }{c \sqrt{ 2\log d_{\mathrm{sh}} } } (d_{\mathrm{sh}})^{-c^2} ( 2\log d_{\mathrm{sh}} )^{ \frac{c^2}{2} } e^{ -\gamma c^2 } \qty( \frac{2}{\pi} )^{ -\frac{c^2}{2} } \nonumber \\
    &= \frac{1}{ c e^{\gamma c^2} } \qty( \frac{\pi}{2} )^{ \frac{c^2-1}{2} } (d_{\mathrm{sh}})^{-(c^2-1)} ( 2\log d_{\mathrm{sh}} )^{ \frac{c^2-1}{2} } \nonumber \\
    &= \frac{1}{ c e^{mc^2} } \qty( \frac{\pi}{2} \frac{1}{d_{0}^2} )^{ \frac{c^2-1}{2} } e^{-(c^2-1)\frac{N}{N_{\mathrm{sh}}} } \qty( \frac{2N}{N_{\mathrm{sh}}} )^{ \frac{c^2-1}{2} } \nonumber \\
    &= \frac{1}{ c e^{\gamma c^2} } \qty( \frac{\pi}{d_{0}^2} \frac{N}{N_{\mathrm{sh}}} )^{ \frac{c^2-1}{2} } e^{-(c^2-1)\frac{N}{N_{\mathrm{sh}}} }. \label{Eq:AppendixA-2}
\end{align}
As noted below Eq.~\eqref{Eq:AppendixA}, this quantity must be small. We see from the last expression in Eq.~\eqref{Eq:AppendixA-2} that this condition is equivalent to $c > 1$.

\clearpage
\section{On the fitting functions and $l$-dependence of $N_{m}$}
\begin{figure}[htb]
    \centering
    \includegraphics[width=0.48\linewidth]{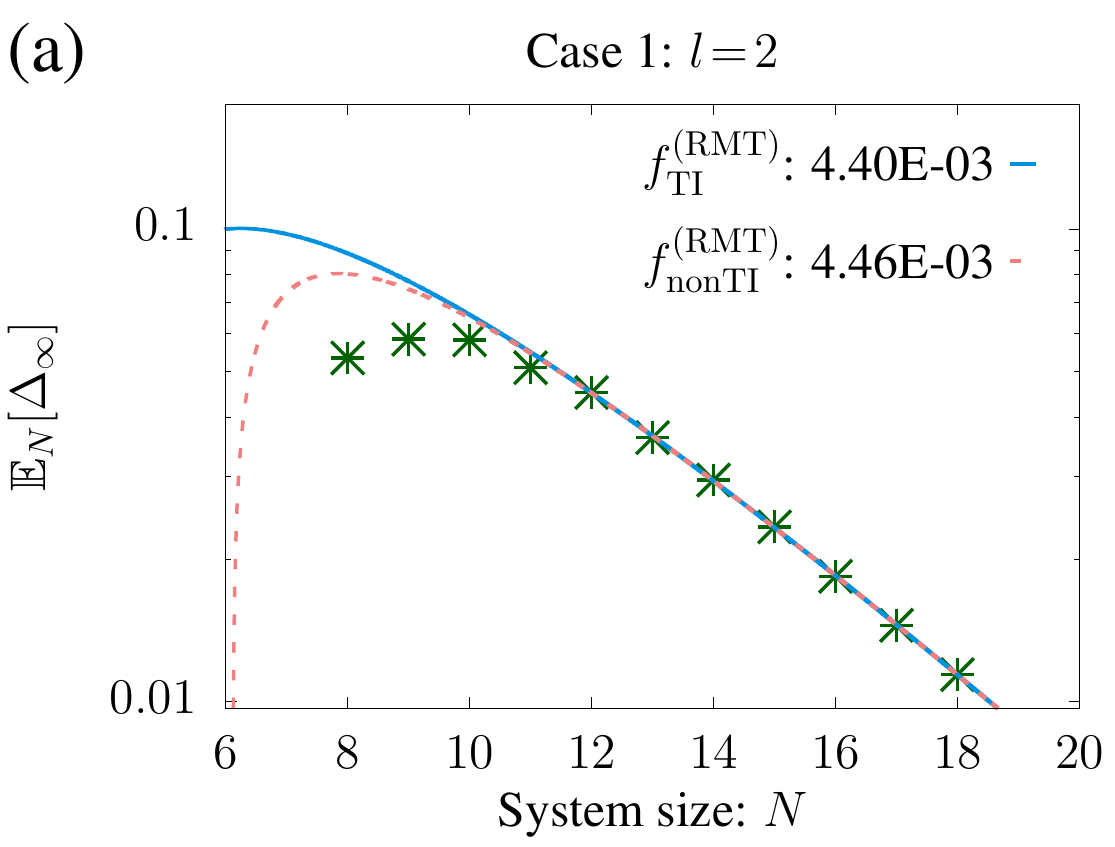}
    \hspace{0.4cm}
    \includegraphics[width=0.48\linewidth]{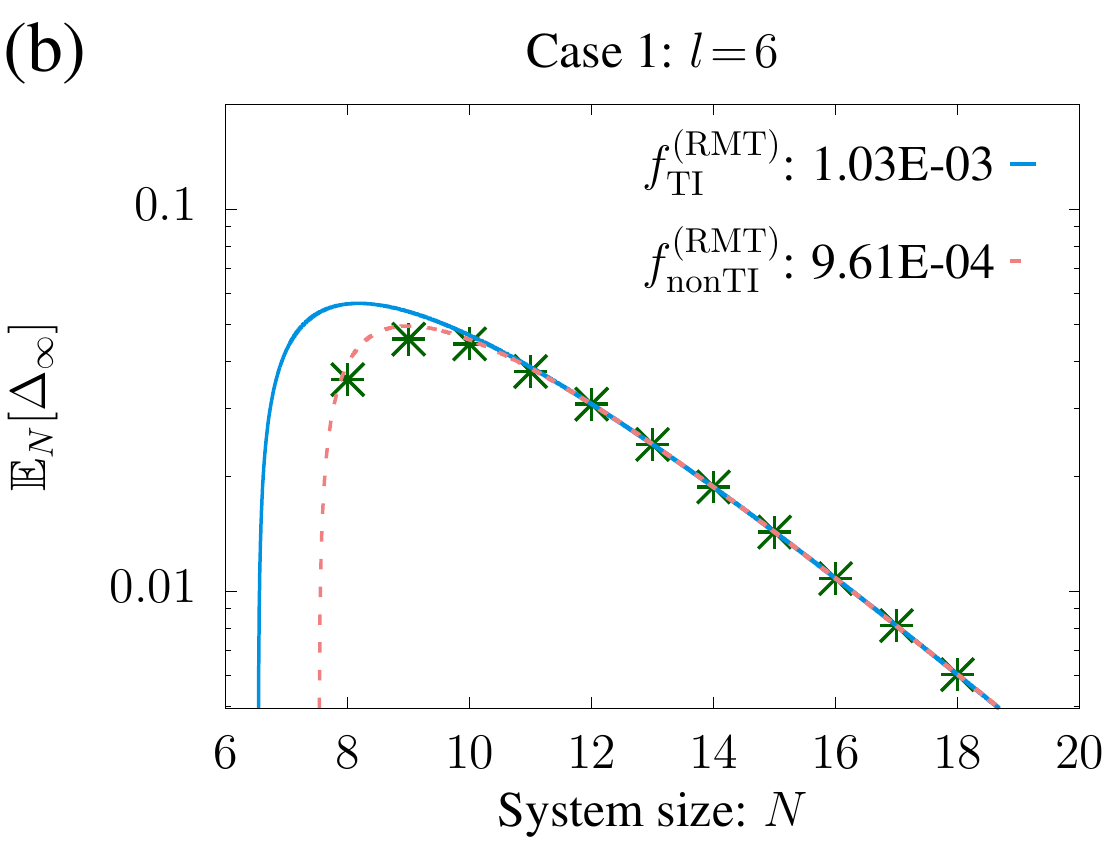}
    \includegraphics[width=0.48\linewidth]{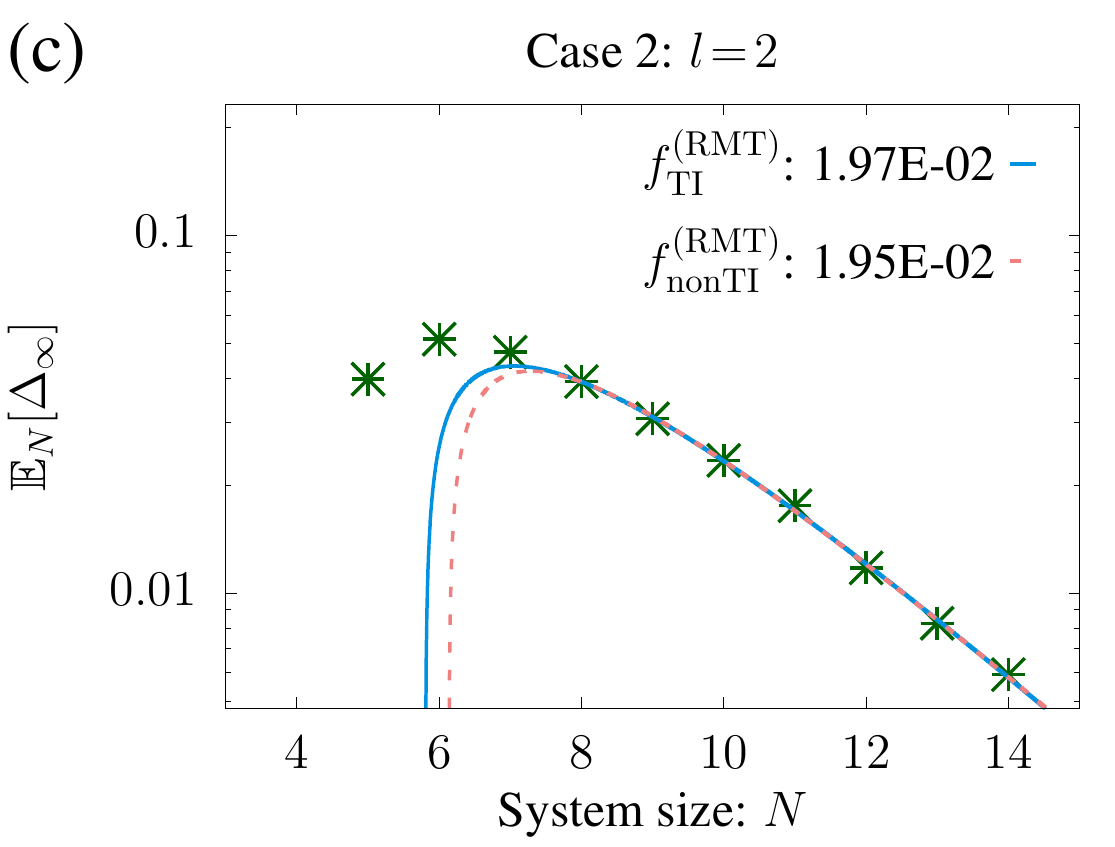}
    \hspace{0.4cm}
    \includegraphics[width=0.48\linewidth]{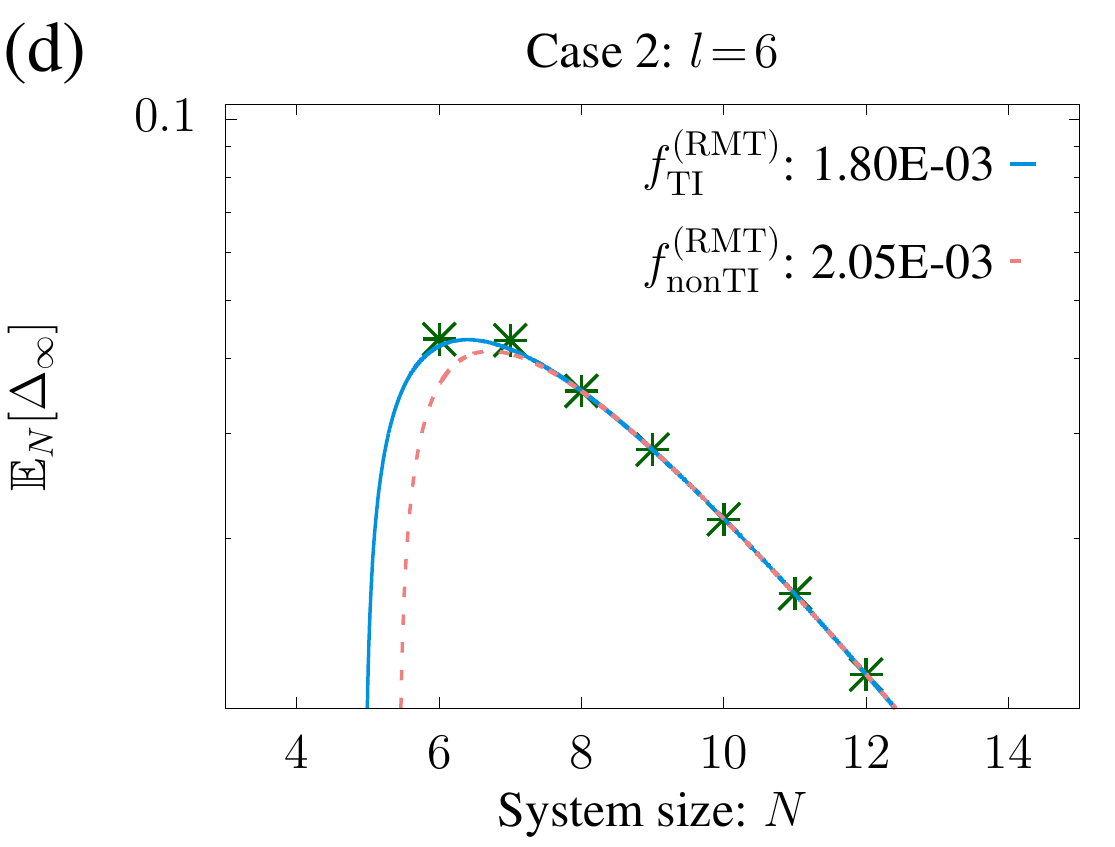}
    \includegraphics[width=0.48\linewidth]{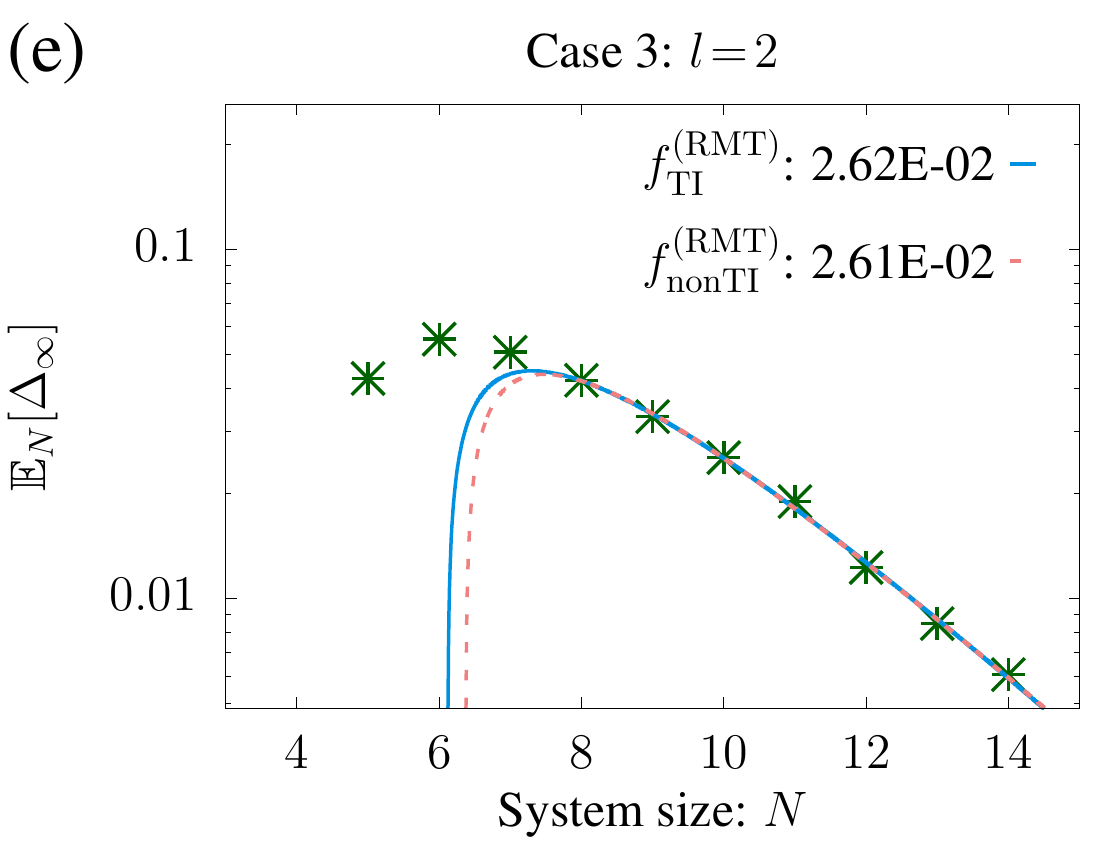}
    \hspace{0.4cm}
    \includegraphics[width=0.48\linewidth]{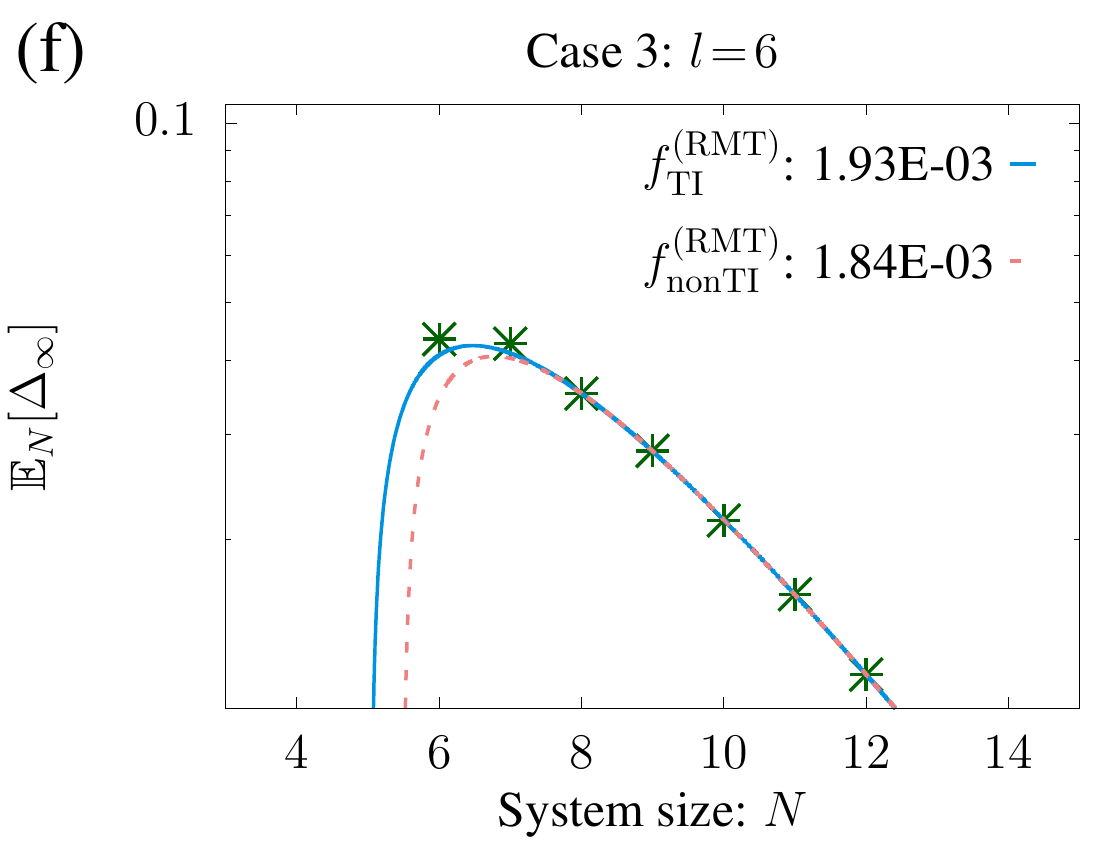}
    \caption{\label{fig:Fittings}
        Results of fitting for ensembles with (a)(c)(e) $l=2$ and (b)(d)(f) $l=6$.
        The functions used here are $f^{(\mathrm{RMT})}_{\mathrm{TI}}$ and $f^{(\mathrm{RMT})}_{\mathrm{nonTI}}$ in Eqs.~\eqref{Eq:Asymptotic_TI} and \eqref{Eq:Asymptotic_nonTI}, respectively, which are derived from the conventional random matrix theory.
        The fitting is carried out in the region $N\geq 12$ for Case~1 and $N\geq 8$ for Cases 2 and 3.
        The numbers at the top right of each panel show the standard deviations of the fitting.
        The deviation in the region with smaller $N$ will be due to a finite-size effect (note that the functions $f^{(\mathrm{RMT})}_{\mathrm{TI}}$ and $f^{(\mathrm{RMT})}_{\mathrm{nonTI}}$ are only asymptotically correct).
    }
\end{figure}

We here discuss the detail of possible fitting functions for $\mathbb{E}_N[\Delta_{\infty}]$.
The asymptotic $N$-dependence of $\mathbb{E}_{N}[\Delta_{\infty}]$ in the conventional random matrix regime can be rigorously obtained as 
\begin{equation}
    f^{(\mathrm{RMT})}_{\mathrm{TI}}(N) = m_{0} N \exp\qty( -\frac{N}{ N_{m} } ) \sqrt{ 1-\frac{ N_{m} }{2} \frac{ \log N }{N} -\frac{ N_{0} }{N} }, \label{Eq:Asymptotic_TI}
\end{equation}
for Case~1 and
\begin{equation}
    f^{(\mathrm{RMT})}_{\mathrm{nonTI}}(N) = m_{0} N^{1/2} \exp\qty( -\frac{N}{ N_{m} } ) \sqrt{ 1 -\frac{ N_{0} }{N} }. \label{Eq:Asymptotic_nonTI}
\end{equation}
for  Case~2 and Case~3 ensembles as in Eqs.~(4) and (5) in the main text (see the main text and the previous section).
In the currently achievable system sizes ($N=18$ for Case~1 and $N=14$ for Case~2 and 3), both Eqs.~(4) and (5) fit equally well to the data in all the cases as shown in Fig.~\ref{fig:Fittings}.

\begin{figure}[htb]
    \centering
    \includegraphics[width=0.48\linewidth]{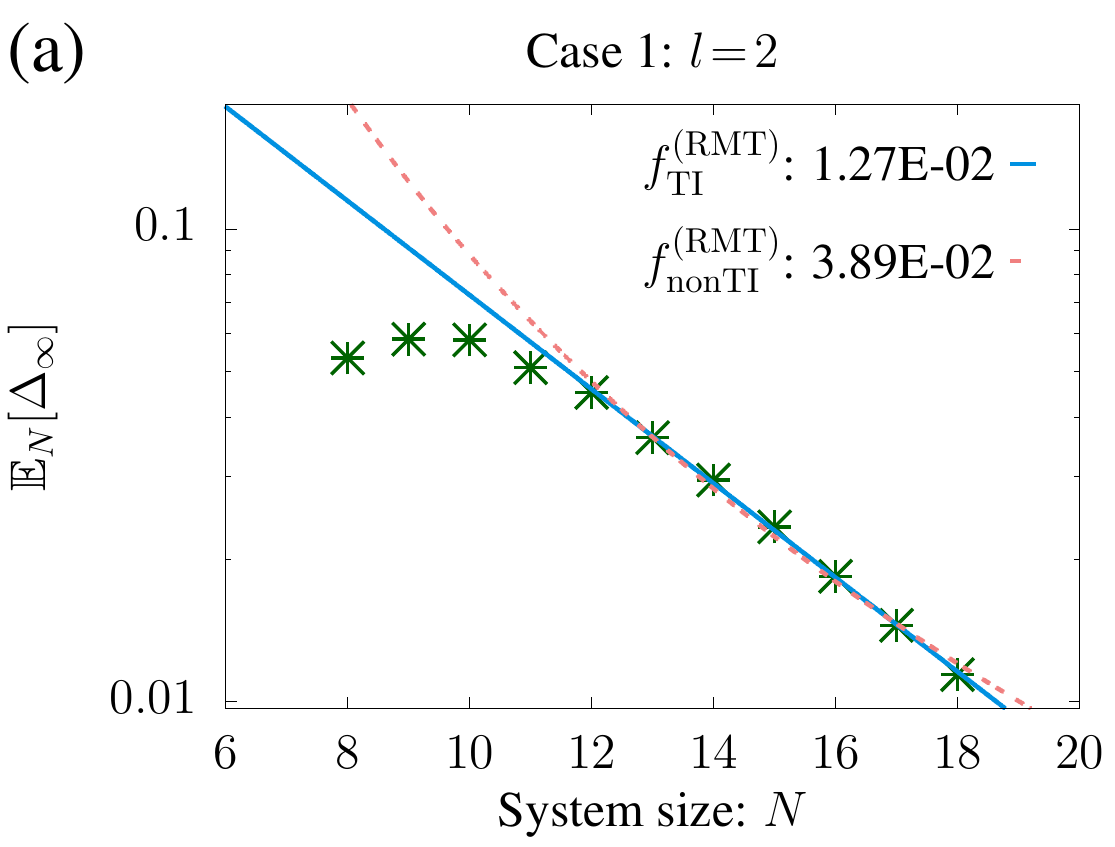}
    \hspace{0.4cm}
    \includegraphics[width=0.48\linewidth]{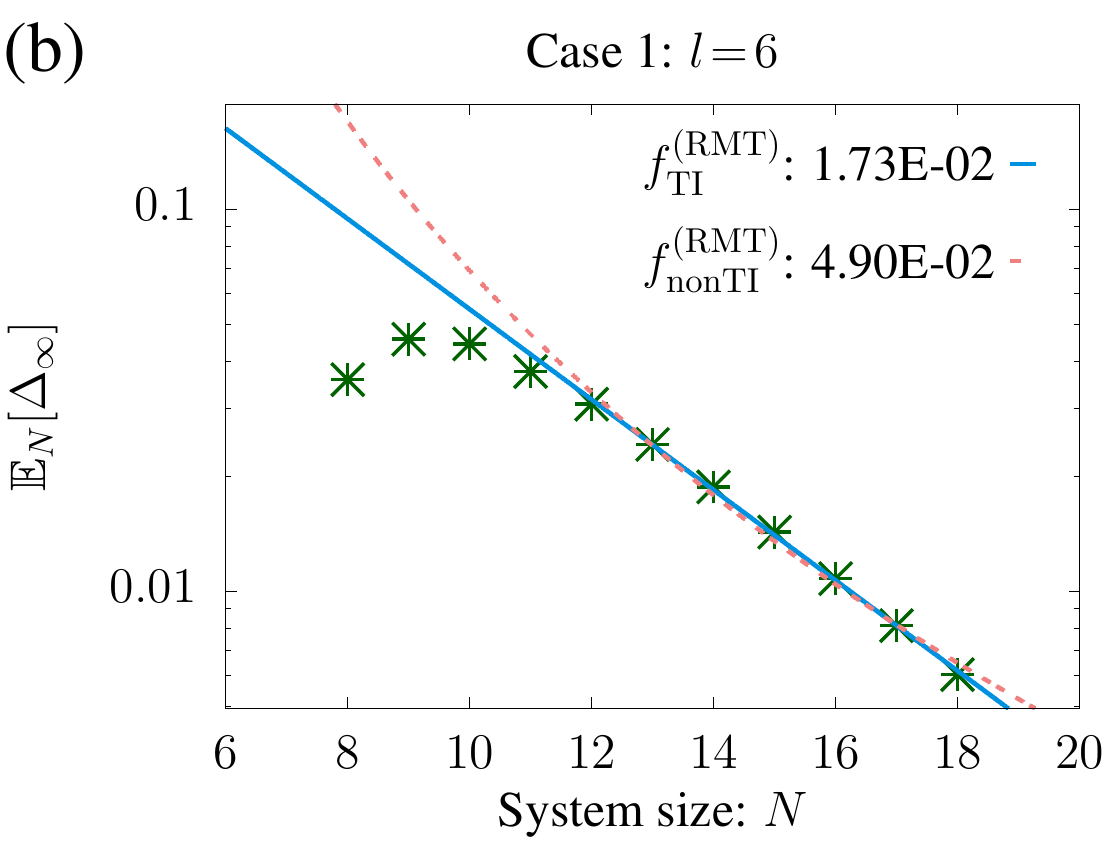}\\
    \includegraphics[width=0.48\linewidth]{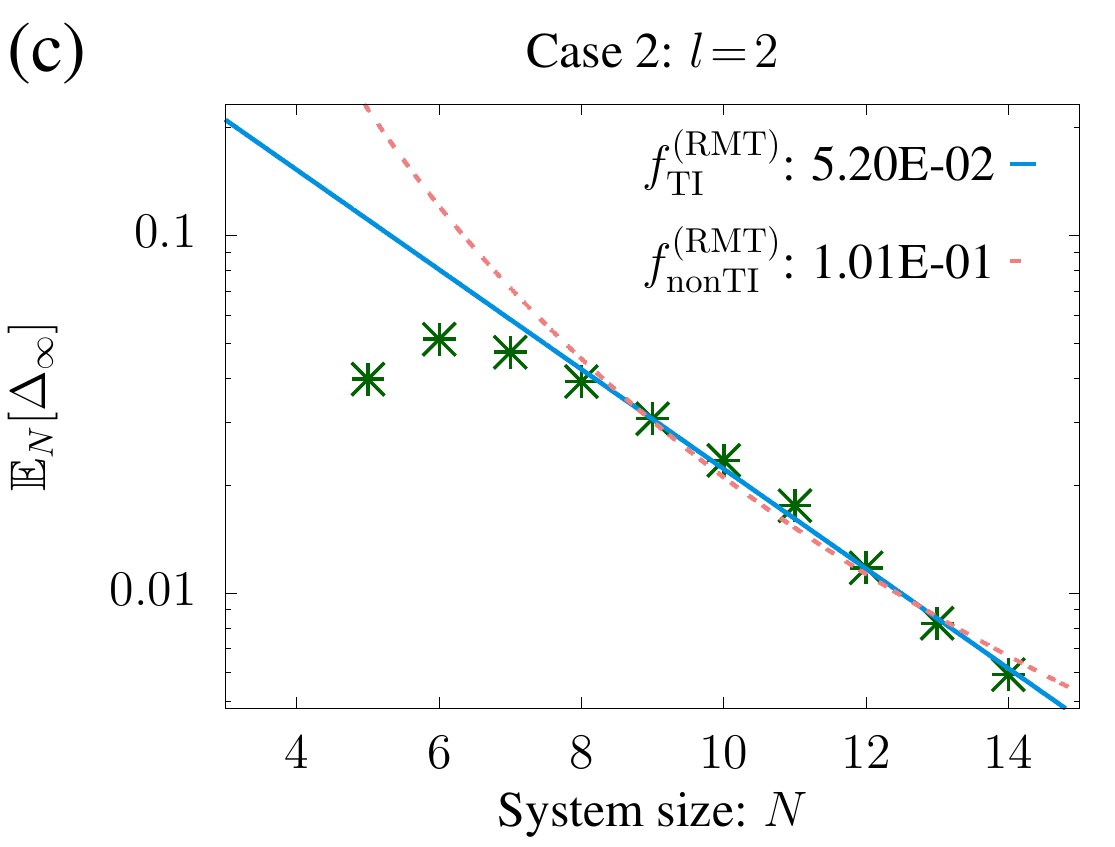}
    \hspace{0.4cm}
    \includegraphics[width=0.48\linewidth]{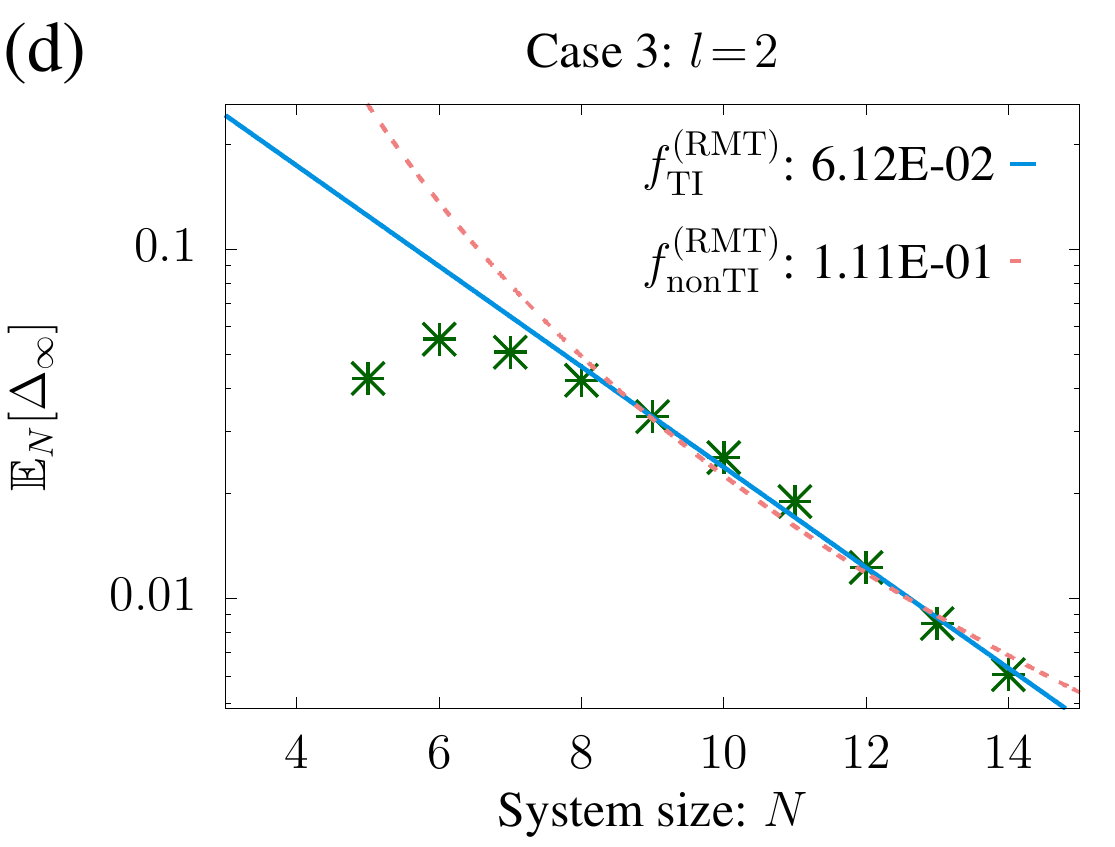}
    \caption{\label{fig:Fittings_Poly}
        Results of fitting to a simple polynomial curve and a simple exponential curve for (a) Case~1 with $l=2$, (b) Case~1 with $l=6$, (c) Case~2 with $l=2$, and (d) Case~3 with $l=2$.
        The numbers at the top right of each panel are the standard deviations of the fitting.
        In all the cases, the exponential fitting is better than the polynomial one.
        Moreover, the polynomial fit deviates from the data with the largest system sizes available ($N=18$ for Case~1 and $N=14$ for Cases 2 and 3).
    }
\end{figure}

In addition, we have compared a simple polynomial fit $(aN^{-b})$ with a simple exponential fit $(ae^{-N/b})$ and find that the polynomial curve does not fit the data compared with the exponential one (Fig.~\ref{fig:Fittings_Poly}).
Therefore, no deviation from the prediction of the conventional random matrix theory (i.e., an exponential decay) is inferred from the data, and we conclude that $\mathbb{E}_{N}\bqty*{ \Delta_{\infty} }$ decreases exponentially as in the conventional random matrix ensemble for large $N$.

\clearpage

\begin{figure}[htb]
    \centering
    \includegraphics[width=0.5\linewidth]{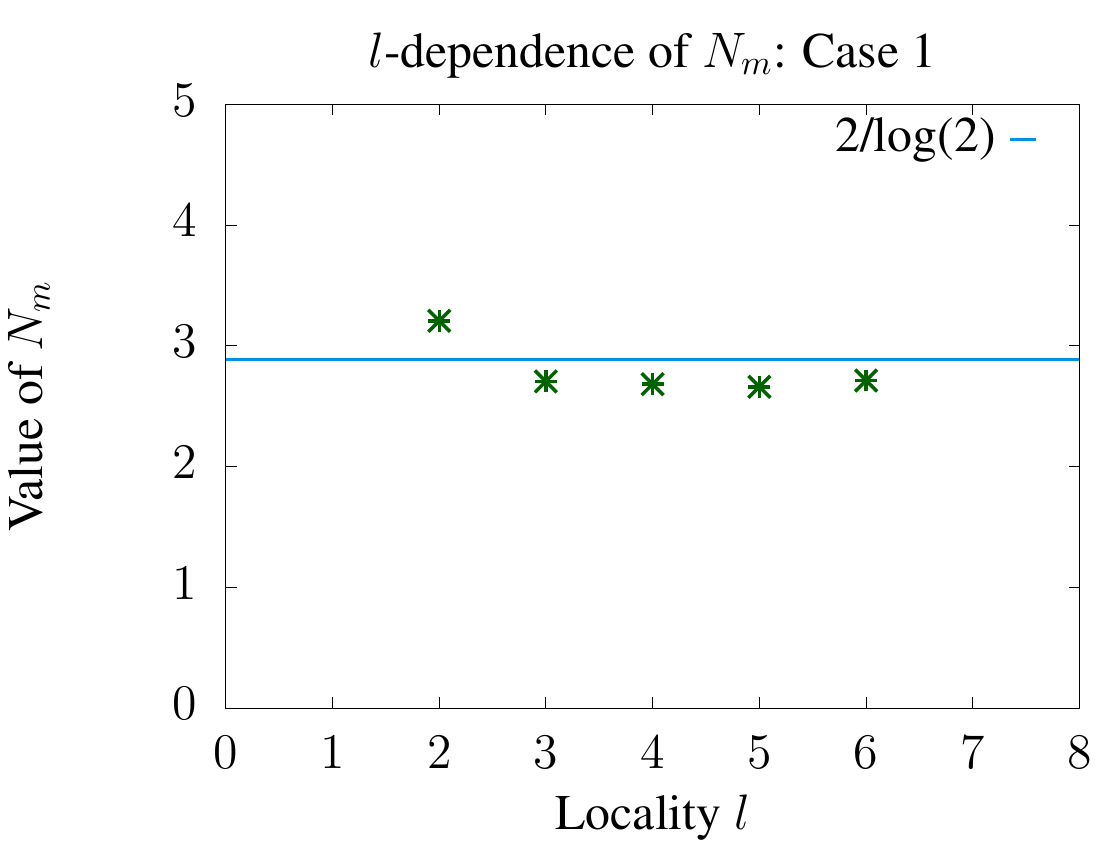}
    \caption{\label{fig:Nmean}
    $l$-dependence of the fitting parameter $N_{m}$ for Case~1.
    The blue line shows the value $2/\log 2$, which is expected from $d_{\mathrm{sh}}\!\sim\! 2^{N}$ and Srednicki's ansatz $s_{N} \sim 1/\sqrt{d_{\mathrm{sh}}}$.
    }
\end{figure}

Next, for these results of fitting, the $l$-dependence of the parameter $N_{m}$ governing the exponential decay does not depend much on $l$ as shown in Fig.~\ref{fig:Nmean}.
This behavior can be understood as follows. 
Given Srednicki's ansatz, the fluctuation $\delta O_{\alpha\alpha}$ and thus $\Delta_\infty$ typically scales as $1/\sqrt{ d_{\mathrm{sh}} }$ (ignoring the logarithmic corrections), where the $N$-dependence of $d_{\mathrm{sh}}$ in the middle of the spectrum does not vary much with $l$.
To see this, we first note that the distribution of $E/\sqrt{\eta_{H}}$ for a single sample converges pointwise to a Gaussian, which can be shown by calculating the moments for $l=\order{1}$ for sufficiently large $N$. 
Thus, if we choose the width $\delta E$ of the shell relative to the spectral range $\eta_{H}$ as $\delta E = \epsilon(N) \eta_{H}$, we have 
\begin{equation}
    d_{\mathrm{sh}} \propto \dim\mathcal{H}_{E,\delta E} \int_{-\epsilon(N) \sqrt{\eta_{H}}}^{+\epsilon(N) \sqrt{\eta_{H}}} \frac{1}{\sqrt{2\pi} \sigma} e^{ -\frac{x^2}{2\sigma^2} } \dd{x}
\end{equation}
with some constant $\sigma$.
On the other hand, we have $d_{\mathrm{sh}} \propto \epsilon(N) \, \dim\mathcal{H}_{N}$ for $l=\order{N}$, since the distribution of $E/\eta_{H}$ converges to Wigner's semi-circle distribution with a constant variance in this case.

Since $\eta_{H} \propto N$ for $l=\order{1}$ for sufficiently large $N$, the difference in the $N$-dependence of $d_{\mathrm{sh}}$ between Case~1 and Case~2, 3 is at most of the order of $\sqrt{N}$, and we expect that it is not as significant as an exponential factor in $N$ (i.e., $\dim\mathcal{H}_N$) even in a region $N\leq 18$.
Therefore, the $l$-dependence of $d_{\mathrm{sh}}$ is determined from that of $\dim\mathcal{H}_N$, and consequently, $N_{m}$ is expected to take an $l$-independent value close to $N_{m}\sim -N/\log(1/\sqrt{d_{\mathrm{sh}}})\sim 2N/\log(\dim\mathcal{H}_N)\sim2/\log d_{\mathrm{loc}}$, which results from the $N$-dependence of the dimension of the total Hilbert space.

In Fig.~\ref{fig:Nmean}, $N_{m}$ takes a value slightly smaller than $2/\log 2$ when $l\geq 3$.
We attribute this discrepancy to an imperfection of the fitting function.
The value of $N_{m}$ for $l=2$ is slightly larger than that for $l=3\sim6$.
This will be because the probability of having atypical samples for which $\Delta_{\infty}$ decreases more slowly than exponential or does not decrease at all is nonnegligible in this case.

\clearpage
\section{Ensembles without translation invariance}
\begin{figure}[tbh]
    \includegraphics[width=0.48\linewidth]{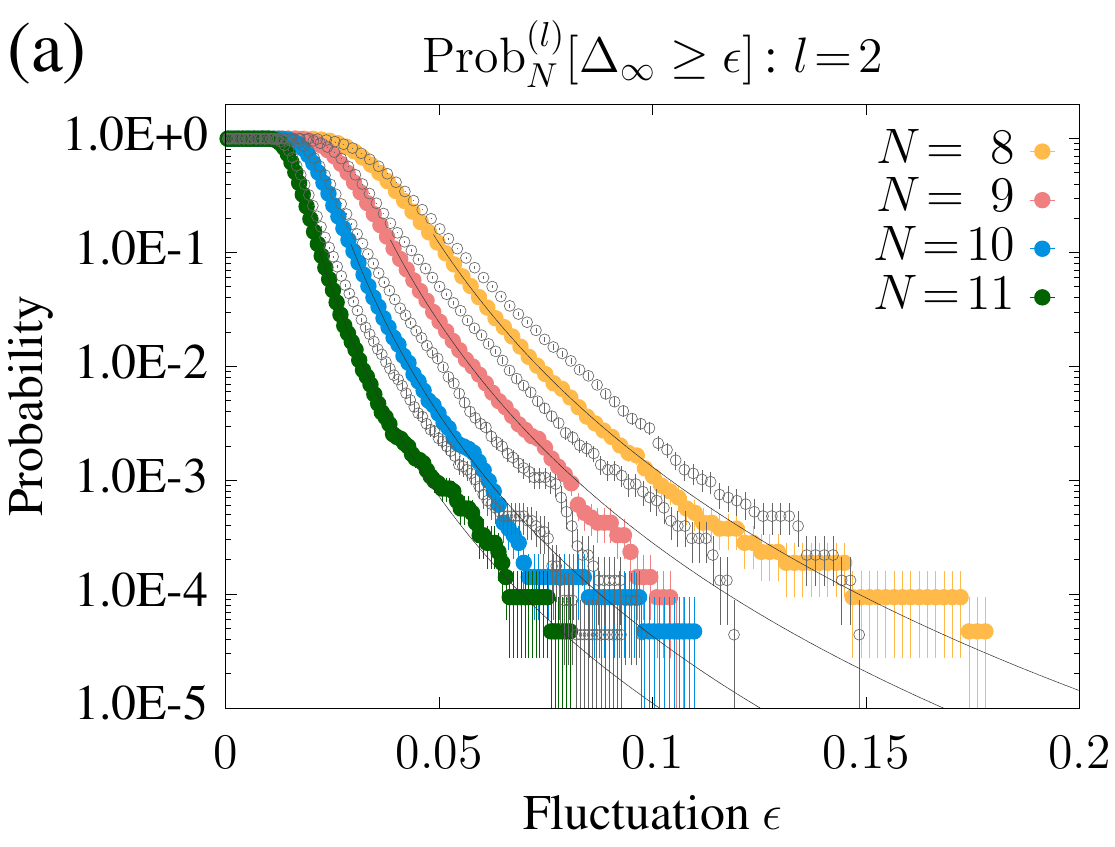}
    \hspace{0.4cm}
    \includegraphics[width=0.48\linewidth]{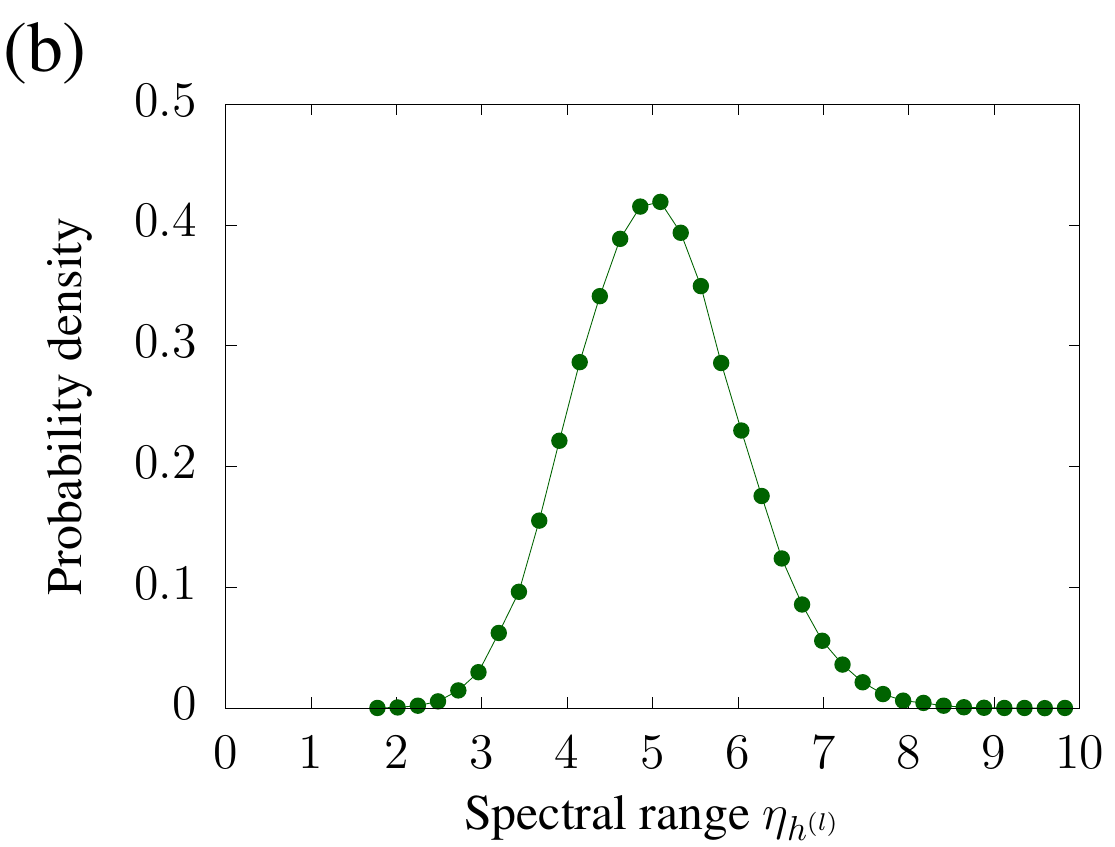}
	\caption{\label{Figure:LocalnonTI_Distribution}
	(a) $\mathrm{Prob}_{N}\bqty*{\Delta_{\infty} \geq \epsilon}$ with $l=2$ for the non-translation-invariant ensembles of Case~2 (colored dots) and Case~3 (gray dots) as a function of $\epsilon$ for various system sizes.
	The dark solid curves are polynomial fits to the data for Case~2.
	The number of the samples is 21220 for Case~2 and 22650 for Case~3.
	(b) Probability density distribution for the spectral range $\eta_{h^{(l)}}$ with $l=2$, where $h^{(l)}$ is sampled from GUE.
	The number of the samples is 100000.
	}
\end{figure}

The distributions of $\Delta_{\infty}$ with $l=2$ for the local but non-translation-invariant ensembles, i.e., Case~2 and Case~3, behave similarly (Fig.~\ref{Figure:LocalnonTI_Distribution}(a)).
Since the only difference between Case~2 and Case~3 is the normalization of $\eta_{h_{j}^{(l)}}$'s, this result implies that the disorder of $\eta_{h_{j}^{(l)}}$ as in Fig.~\ref{Figure:LocalnonTI_Distribution}(b) does not prevent thermalization.

The mean of the distribution of $\Delta_{\infty}$ with $l=2$ is smaller for the non-translation-invariant ensembles of Case~2 and Case~3 than the translation invariant ensemble of Case~1.
Thus, we find that the randomness of an interaction in a single Hamiltonian favorably contributes to the strong ETH in most cases as long as the strength of interactions does not change too much.

\clearpage
\section{Srednicki's conjecture}
Here, we investigate the validity of Srednicki's conjecture~\cite{srednicki1999approach} for our local random matrix ensembles.
The fluctuations of the eigenstate expectation values are defined as
$
    \delta(O_{N})_{\alpha\alpha} \coloneqq (O_{N})_{\alpha\alpha} -\expval*{ \hat{O}_{N} }^{\mathrm{mc}}_{ E_{\alpha},\delta E }.
$
We decompose it into two terms:
\begin{align}
    \delta (O_{N})_{\alpha\alpha} &= \mathcal{E}_{\gamma}^{(E_{\alpha},\delta E)}[ \delta(O_{N})_{\gamma\gamma} ] +\mathcal{S}_{\gamma}^{(E_{\alpha},\delta E)}[ \delta(O_{N})_{\gamma\gamma} ] \tilde{R}_{\alpha\alpha}
\end{align}
where $\tilde{R}_{\alpha\alpha}$ is a pseudo-random variable over the index $\alpha$ of the eigenstates. 
Here, $\mathcal{E}_{\gamma}^{(E_{\alpha},\delta E)}$ and $\mathcal{S}_{\gamma}^{(E_{\alpha},\delta E)}$ represent the spectral average and the standard deviation within an energy shell $\mathcal{H}_{E_{\alpha}, \delta E}$.
The ETH asserts that both $\mathcal{E}_{\gamma}^{(E_{\alpha},\delta E)}[ \delta(O_{N})_{\gamma\gamma} ]$ and $\mathcal{S}_{\gamma}^{(E_{\alpha},\delta E)}[ \delta(O_{N})_{\gamma\gamma} ]$ vanish in the thermodynamic limit.

\begin{figure}[tb]
    \centering
    \includegraphics[width=0.48\linewidth]{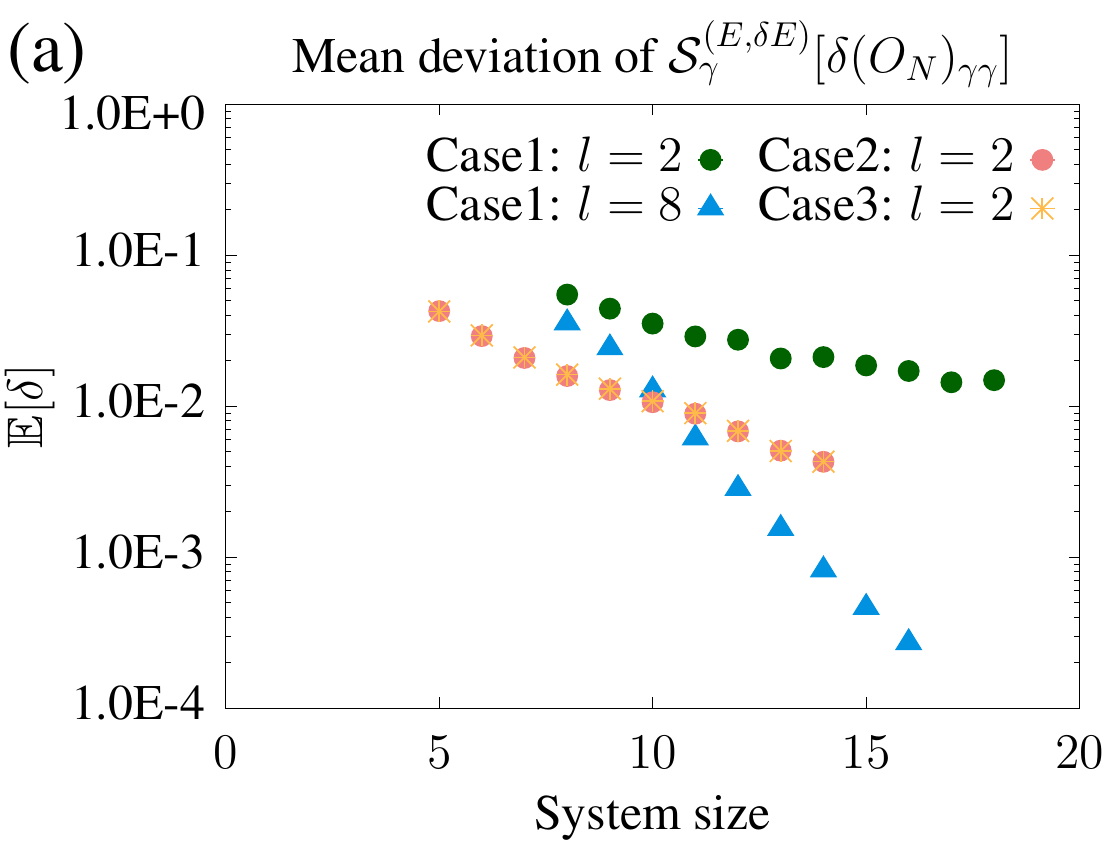}
    \hspace{0.4cm}
    \includegraphics[width=0.48\linewidth]{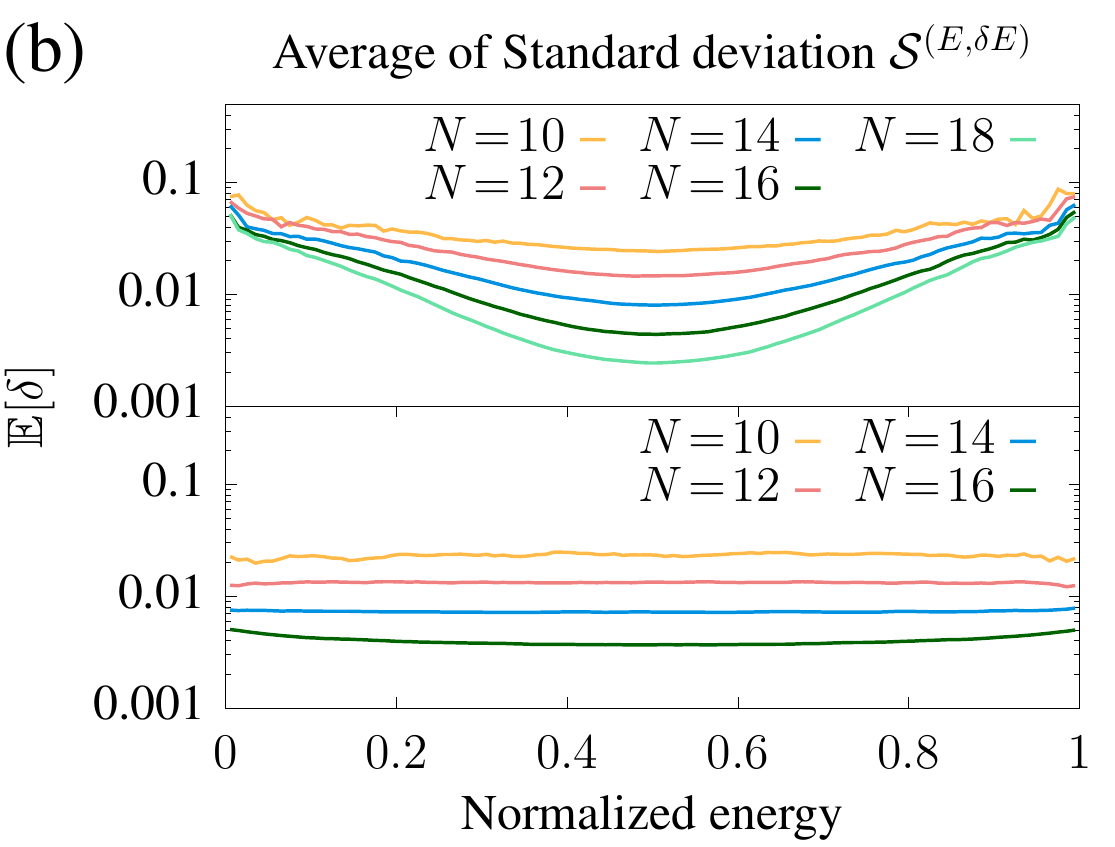}
    \caption{ \label{Figure:Stddev}
    (a) Mean deviation of $\mathcal{S}^{(E, \delta E)}_{\gamma}[ \delta(O_{N})_{\gamma\gamma} ]$ from its ensemble average. It decreases in all cases, which implies the weak ETH.
    (b) Ensemble average of $\mathcal{S}^{(E, \delta E)}_{\gamma}[ \delta(O_{N})_{\gamma\gamma} ]$ for Case~1. 
    For the local ensemble with $l=2$ (up), it depends on the energy eigenvalue and decreases with increasing $N$ in the middle of the spectrum but does not decrease so much near the edges. 
    On the other hand, for the global ensemble with $l=8$ (down), it does not depend on energy and uniformly decreases with increasing $N$ in the spectrum. 
    The number of samples is 1022 for Case~1 with $l=2$, 379 for Case~1 with $l=8$, and 10000 for Case~2 and Case~3.
    }
\end{figure}

The mean $L$2-norm of the deviation from the ensemble average of the standard deviation of the eigenstate fluctuation in an energy shell typically decreases with the system size for all of our ensembles (Fig.~\ref{Figure:Stddev} (a)).
In addition, $\mathcal{S}_{\gamma}^{(E_{\alpha},\delta E)}[ \delta(O_{N})_{\gamma\gamma} ]$ itself decreases in the middle of the spectrum irrespective of the locality. 
This fact reflects the weak ETH, which is shown rigorously for systems with translation invariance and exponential decay of correlation functions. 
However, if we impose the locality of interactions, the magnitude of eigenstate fluctuations $\mathcal{S}_{\gamma}^{(E_{\alpha},\delta E)}[ \delta(O_{N})_{\gamma\gamma} ]$ acquires the energy dependence, and it does not decrease so much near the edges (Fig.~\ref{Figure:Stddev} (b)). 
This result is consistent with the previous study~\cite{brandino2012quench}, where the locality of interactions is imposed as the sparsity of the Hamiltonian matrix.

\begin{figure}[htb]
    \centering
    \includegraphics[width=0.48\linewidth]{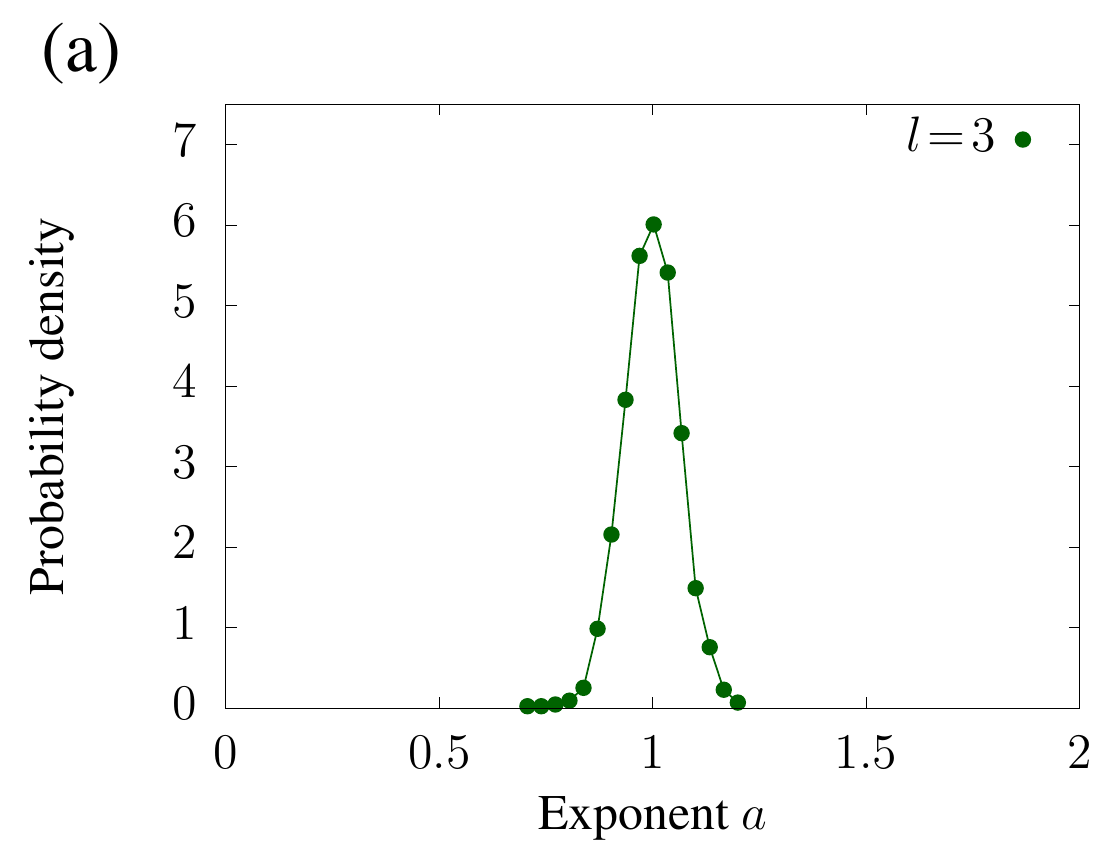}
    \hspace{0.4cm}
    \includegraphics[width=0.48\linewidth]{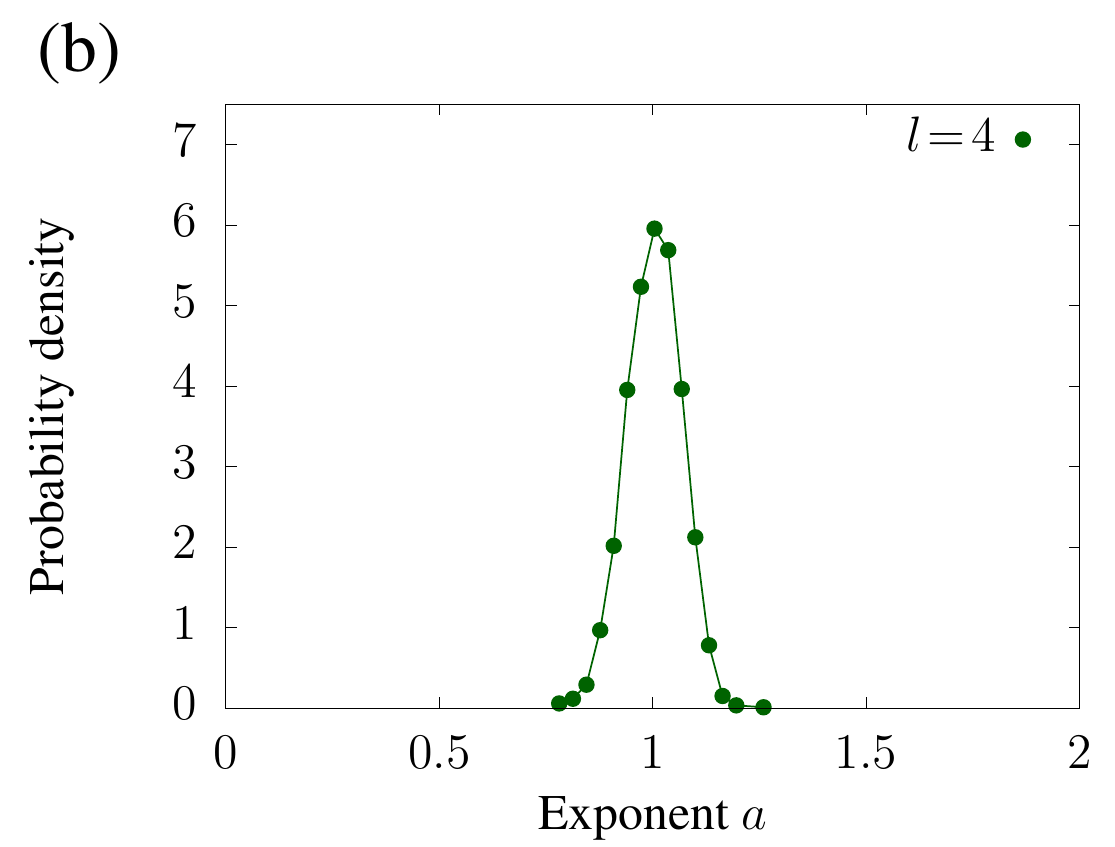}
    \caption{\label{fig:StddevInShell_Exponent_Histogram}
    Probability distribution of $a$ that fits $\mathcal{S}_{\gamma}^{(E,\delta E)} \propto (\sqrt{ d_{\mathrm{sh}} })^{-a}$ for Case~1 with (a)~$l=3$ and (b)~$l=4$, where $E$ is chosen to be at the center of the spectrum and $\delta E$ is 5\% of the spectral range.
    The distribution peaks around $a=1$ irrespective of the locality $l$.
    The number of the samples is (a)~1325 and (b)~2688.
    }

    \vspace{1cm}
    \includegraphics[width=0.48\linewidth]{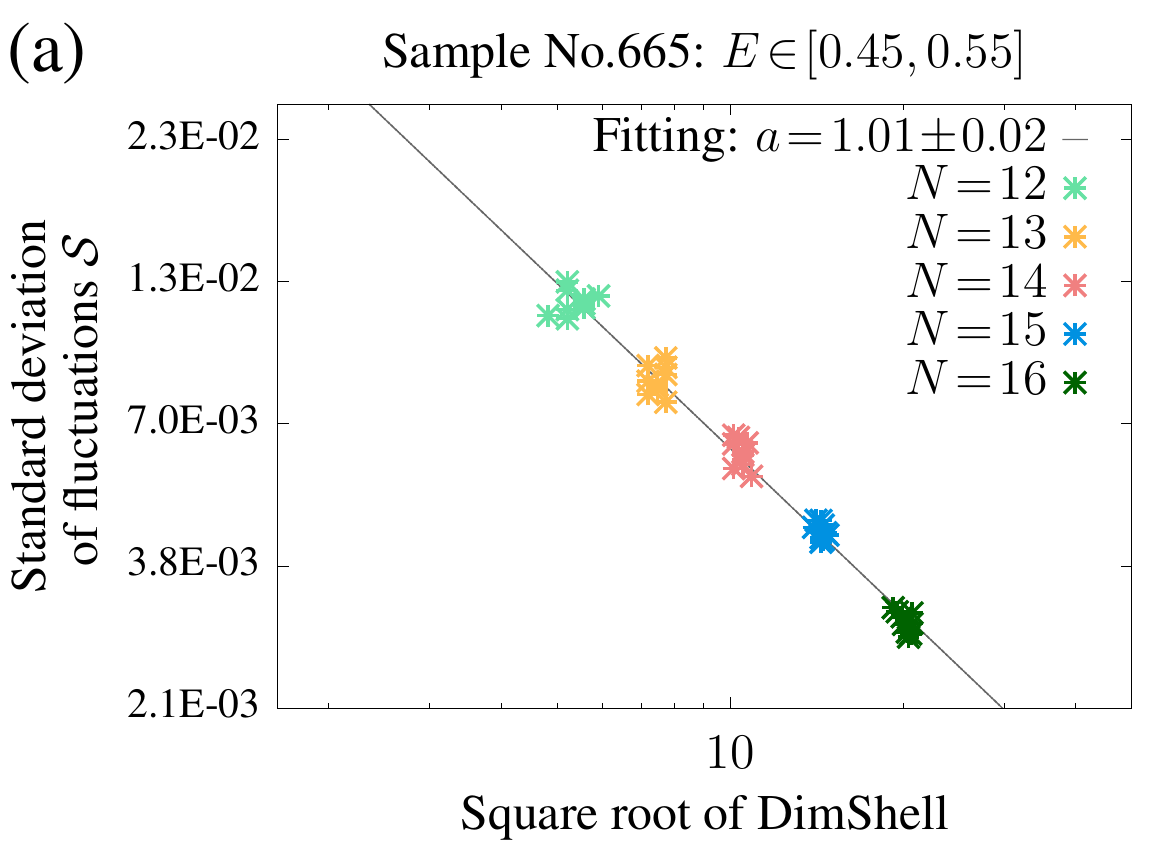}
    \hspace{0.4cm}
    \includegraphics[width=0.48\linewidth]{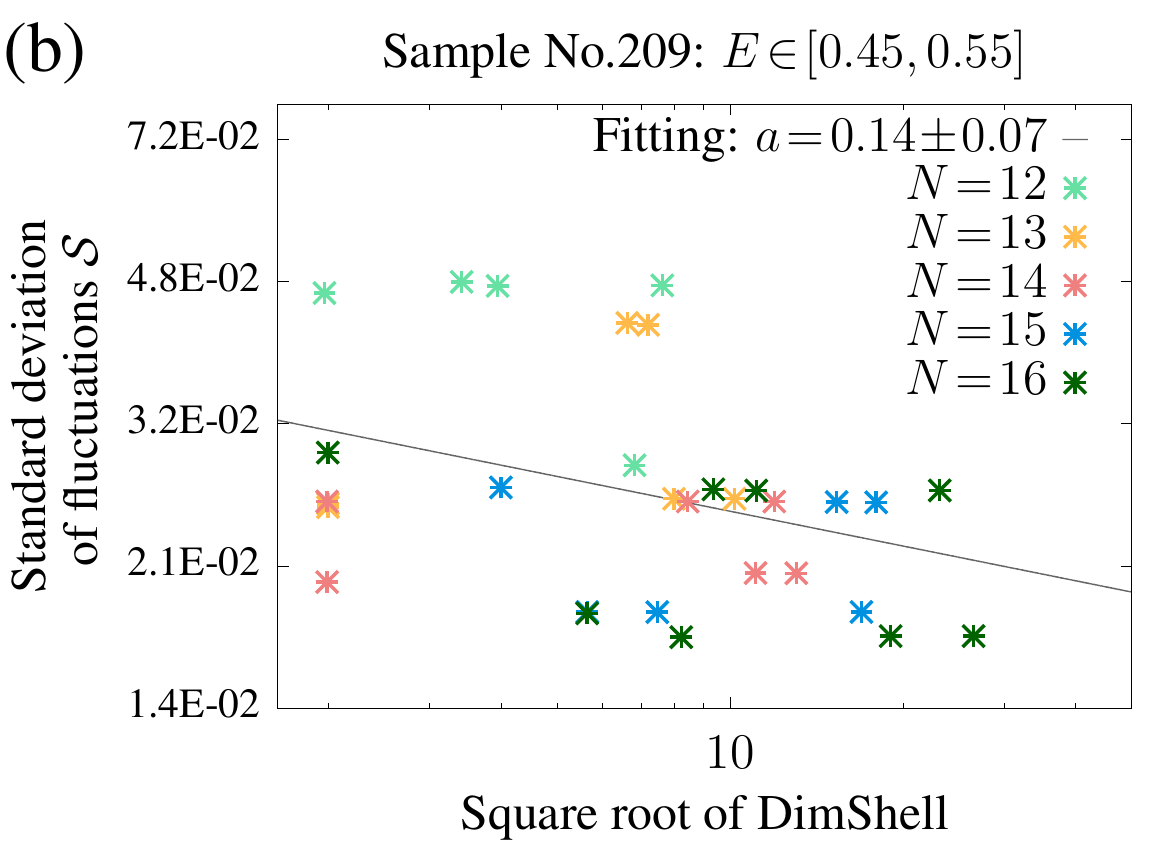}
    \caption{\label{fig:Stddev_vsDimShell}
    $\sqrt{d_{\mathrm{sh}}}$-dependence of $\mathcal{S}^{(E,\delta E)}_{\gamma}$ for (a) a typical sample of $\hat{h}^{(2)}$ with $a=1.01$ and (b) an atypical one with $a=0.14$ both taken from  Case~1 ensemble with $l=2$.
    For the sample with $a=0.14$, the standard deviation $\mathcal{S}_{\gamma}^{(E,\delta E)}$ decreases neither with $d_{\mathrm{sh}}$ nor with $N$.
    }
\end{figure}

\begin{figure}[htb]
    \centering
    \includegraphics[width=0.48\linewidth]{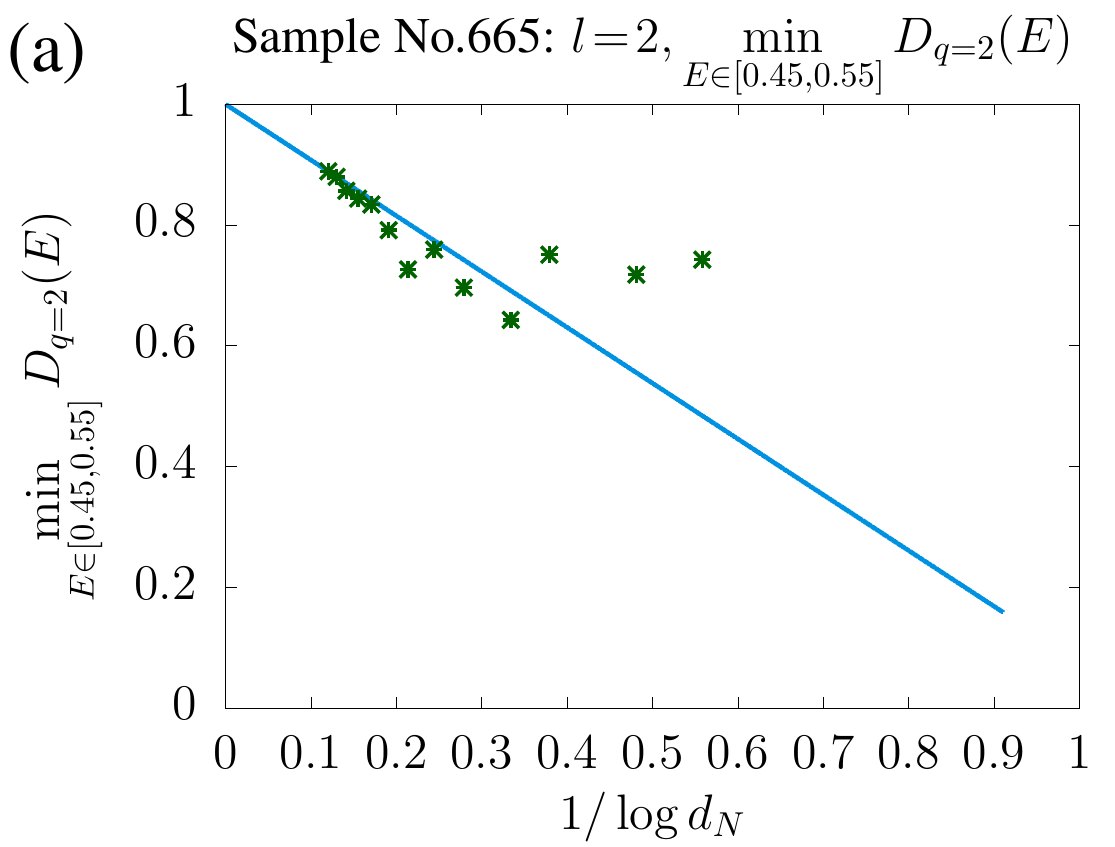}
    \hspace{0.4cm}
    \includegraphics[width=0.48\linewidth]{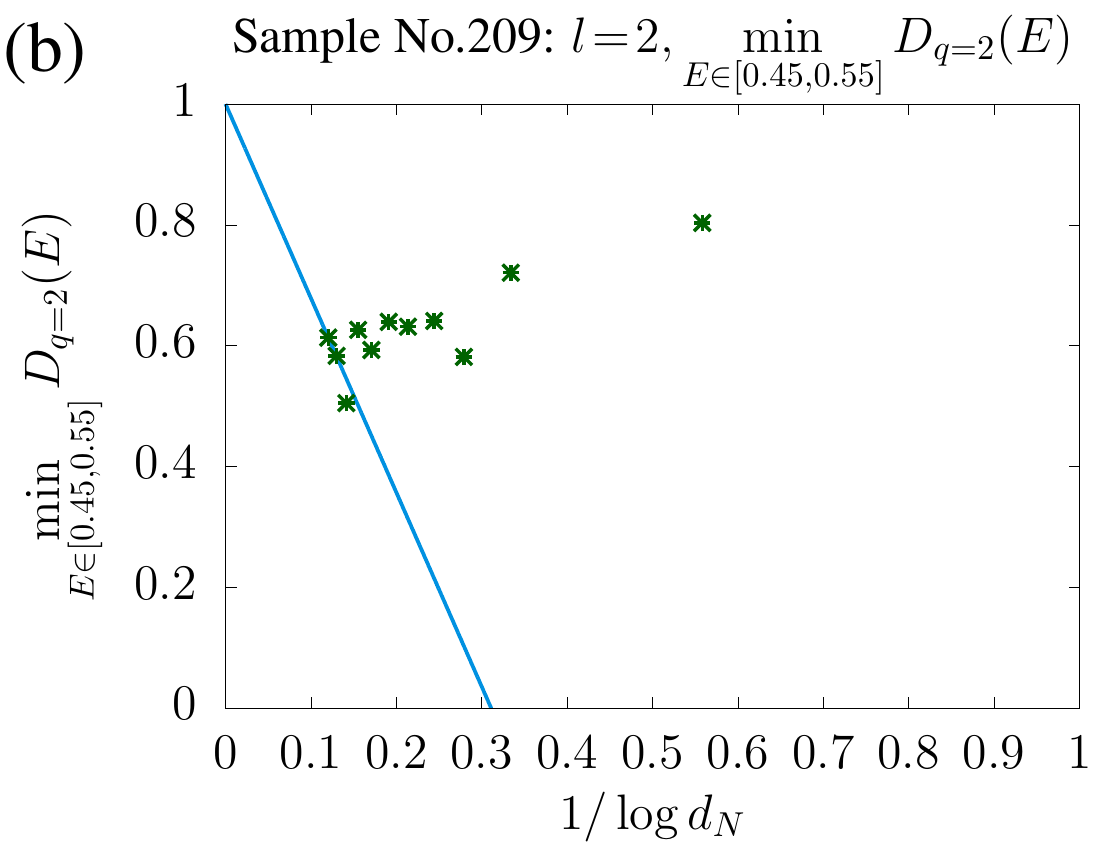}
    \caption{\label{fig:MultiFractal_vsDimtot}
    Minimum multi-fractal dimension among energy eigenstates in the middle 10\% of the spectrum for (a) a typical sample of $\hat{h}^{(2)}$ with $a=1.01$ and (b) an atypical one with $a=0.14$.
    The blue line is a guide to the eye that connects the data point with the largest $d_{N}$ and $(0,1)$.
    The data points indicate that for typical Hamiltonians with $a\simeq 1$ the multi-fractal dimension tends to converge to 1 as $N\to\infty$, but that for atypical Hamiltonians there remain eigenstates with a multi-fractal dimension less than one.
    In this sense, atypical Hamiltoninas have multi-fractal eigenstates.
    }
    \vspace{1cm}
    \includegraphics[width=0.48\linewidth]{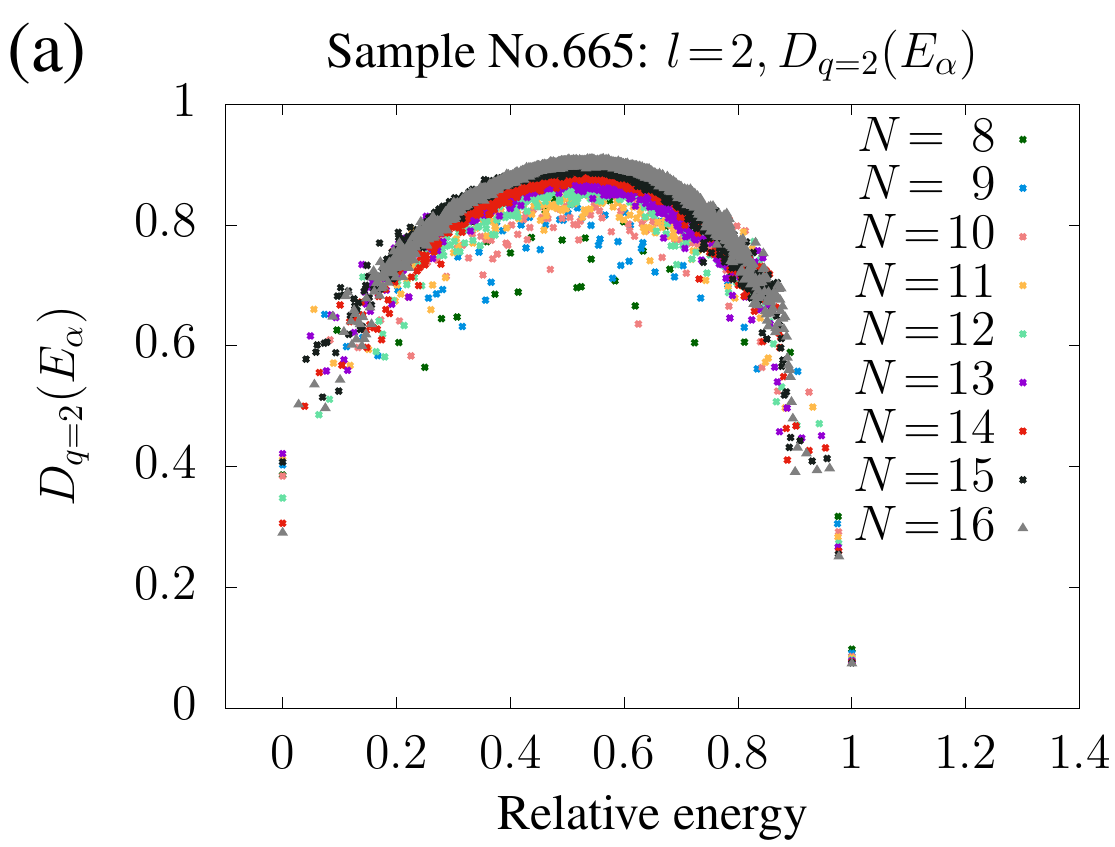}
    \hspace{0.4cm}
    \includegraphics[width=0.48\linewidth]{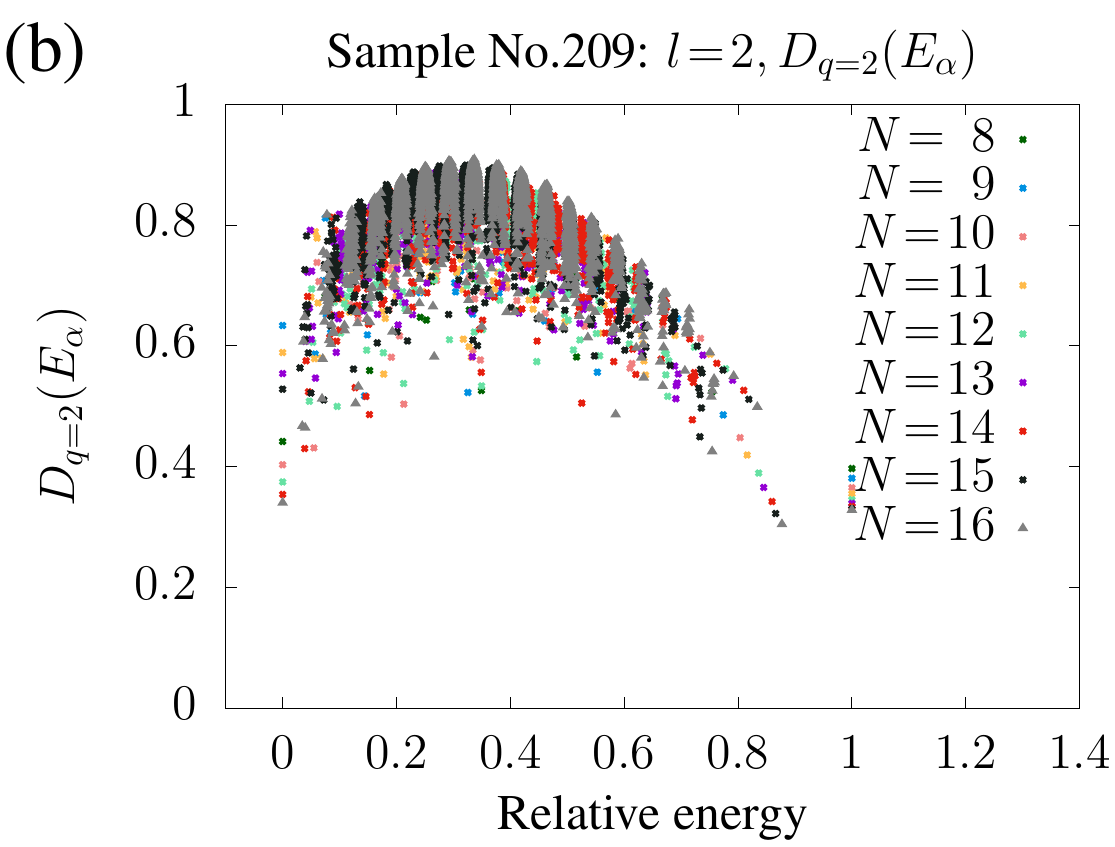}
    \caption{\label{fig:MultiFractal_vsEnergy}
    Multi-fractal dimension of energy eigenstates for (a) a typical sample of $\hat{h}^{(2)}$ with $a=1.01$ and (b) an atypical one with $a=0.14$.
    }
\end{figure}

As shown in FIG.~3 in the main text, we find that Srednicki's ansatz that $\mathcal{S}^{(E, \delta E)}_{\gamma} \propto (\sqrt{ d_{\mathrm{sh}} })^{-a}$ with $a\simeq 1$ holds for typical samples irrespective of  locality $l$.
We note that the distribution of $a$ for $l=3,4$ also peaks around $a=1$ (Fig.~\ref{fig:StddevInShell_Exponent_Histogram}).
However, for the case with $l=2$, atypical samples that have small $a$ exist as demonstrated in Fig.~\ref{fig:Stddev_vsDimShell}, and there appear multi-fractal eigenstates even in the middle of the spectrum for those samples in the sense that the minimum of (finite) multi-fractal dimension defined by
\begin{equation}
    D_{q}(E_{\alpha}) = -\frac{1}{q-1} \frac{1}{\log d_{N}} \log\qty( \sum_{j=1}^{d_{N}} \abs*{ c_{j}^{(\alpha)} }^{2q} )\qc c_{j}^{(\alpha)} = \braket{j}{E_{\alpha}}\qc \ket*{j} \in \mathcal{B}_{N}, \label{Eq:MultiFractal}
\end{equation}
(especially for $q=2$) does not decrease in the thermodynamics limit (Fig.~\ref{fig:MultiFractal_vsDimtot})~\cite{backer2019multifractal}.
Figure~\ref{fig:MultiFractal_vsEnergy} shows $D_{q=2}(E_{\alpha})$ of eigenstates $\ket*{E_{\alpha}}$ with respect to an eigenbasis $\{\ket{j}\}$ of the translation operator $\hat{T}_{N}$ for two realizations of the Case~1 ensemble, where $d_{N}\coloneqq d_{\mathrm{loc}}^{\ N} / N$ is the dimension of the total Hilbert space (i.e., the zero-momentum sector in our calculation).
This clearly shows that eigenstates with a small multifractal dimension appear more frequently for atypical samples than for typical ones.
We expect that these atypical Hamiltonians are also responsible for the long tail of the distribution of $\Delta_{\infty}$.

Srednicki's conjecture also asserts that $\tilde{R}_{\alpha\alpha}$'s for a generic single Hamiltonian $\hat{H}_{N}$ distribute according to a normal Gaussian distribution and $\mathcal{S}_{\gamma}^{(E,\delta E)}[ \delta(O_{N})_{\gamma\gamma} ]$ can be decomposed as $\mathcal{S}_{\gamma}^{(E,\delta E)}[ \delta(O_{N})_{\gamma\gamma} ] = e^{ -S(E)/2 } f_{O}(E)$, where $S(E)$ is the thermodynamic entropy of the system and $f_{O}$ represents an energy dependence of the magnitude of eigenstate fluctuations. Although the exact formula to calculate the thermodynamic entropy from the microscopic description of the system has not been obtained yet~\footnote{Boltzmann's formula contains a subextensive ambiguity concerning the width of an energy shell $\delta E$. }, the system-size dependence of the magnitude of eigenstate fluctuations arises from the factor $e^{ -S(E)/2 }$.

\begin{figure}[tb]
    \centering
    \includegraphics[width=0.48\linewidth]{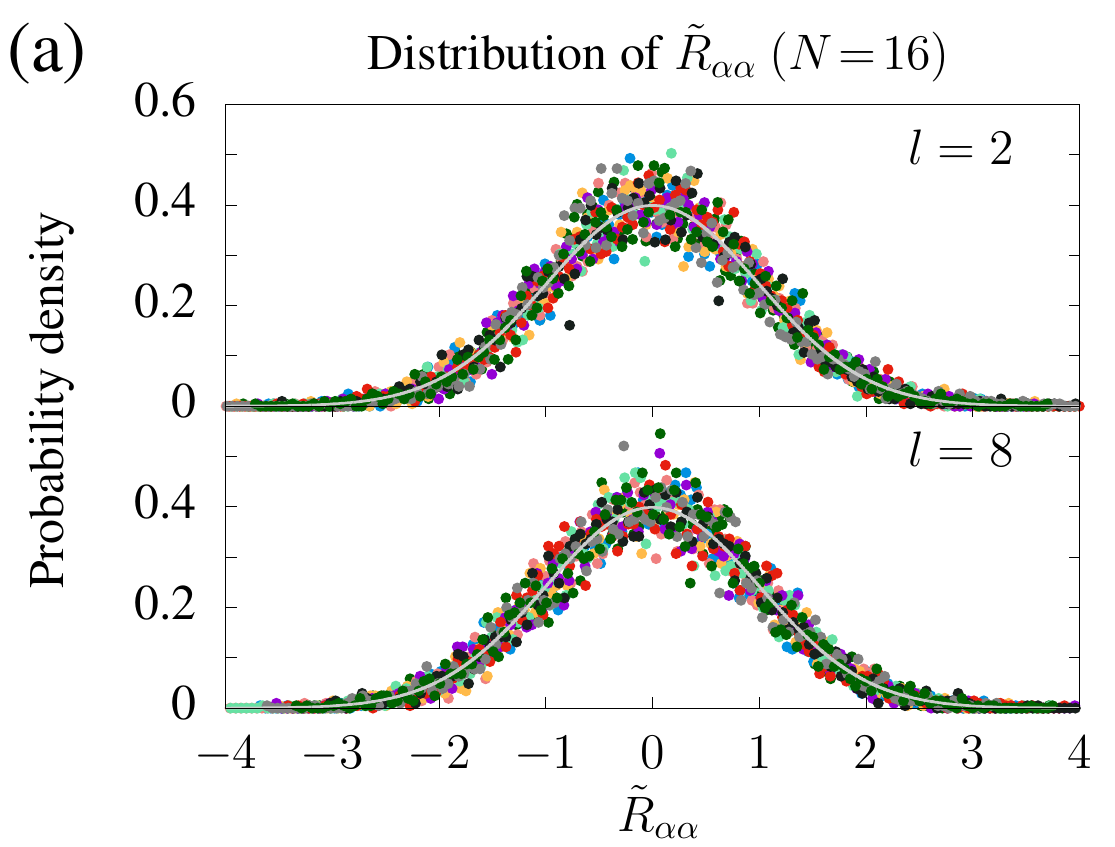}
    \hspace{0.4cm}
    \includegraphics[width=0.48\linewidth]{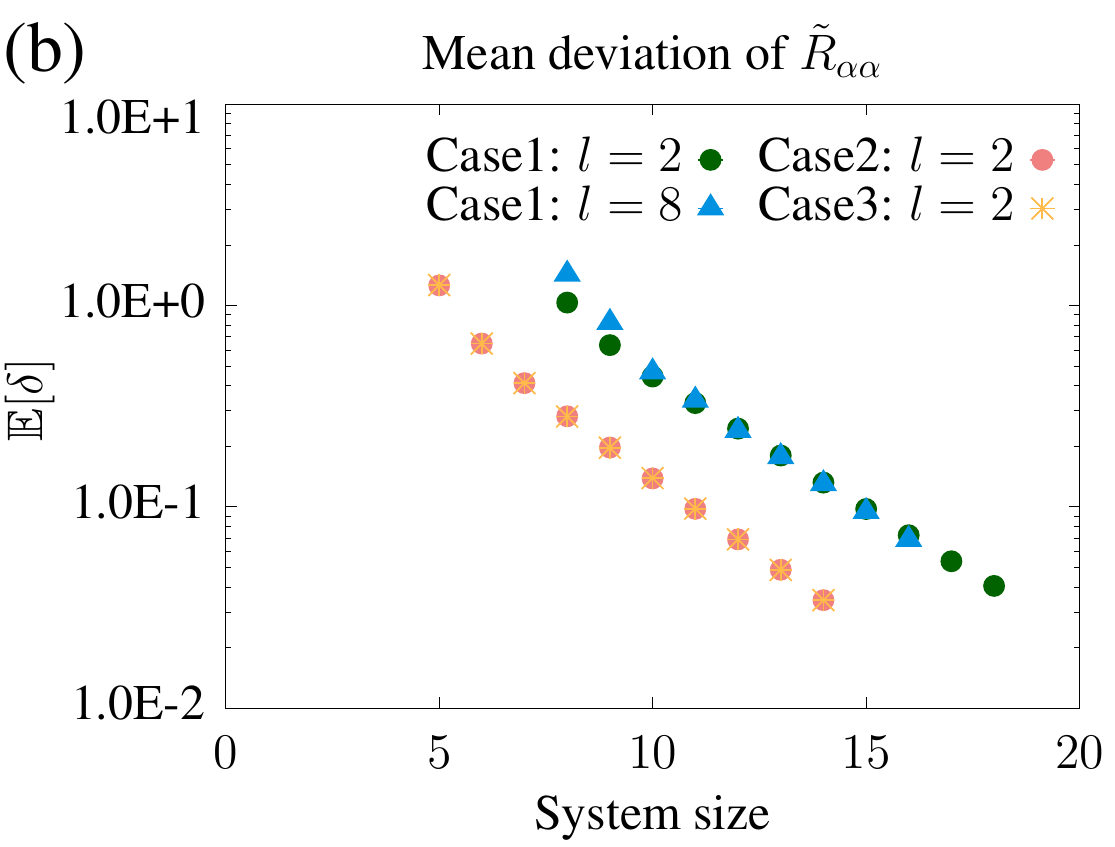}
    \caption{\label{Figure:SrednickiAnsaz}
    (a) Distribution of the pseudo-random variable $\tilde{R}_{\alpha\alpha}$ over the entire spectrum for 10 realizations of Hamiltonians taken from the Case~1 ensemble with $l=2$ (upper panel) and that with $l=8$ (lower panel). 
    The gray curves show the normal Gaussian distributions. 
    In both cases, the distributions agree excellently with the normal distribution, which is consistent with Srednicki's ansatz.
    (b) Mean deviation of the distribution of $\tilde{R}_{\alpha\alpha}$ from the normal Gaussian distribution. 
    Its magnitude decreases exponentially in all cases.
    Thus, we find that the pseudo-random variable $\tilde{R}_{\alpha\alpha}$ still distributes according to the normal Gaussian distribution even in the presence of the locality of interactions.
    The number of samples is 1022 for Case~1 with $l=2$, 379 for Case~1 with $l=8$, and 10000 for Case~2 and Case~3.
    }
\end{figure}

We here show the distributions of $\tilde{R}_{\alpha\alpha}$ for 10 realizations of random Hamiltonians taken from a local ensemble, i.e., Case~1 with $l=2$ (Fig.~\ref{Figure:SrednickiAnsaz}(a), up), and a global ensemble, i.e., Case~1 with $l=8$ (Fig.~\ref{Figure:SrednickiAnsaz}(a), down). 
We also calculate the mean deviation of the distributions of $\tilde{R}_{\alpha\alpha}$ from the normal Gaussian for all of our ensembles, and find that they decrease in almost the same way irrespective of the locality (Fig.~\ref{Figure:SrednickiAnsaz}(b)).
These results imply that the locality of interactions does not affect the validity of Srednicki's conjecture.

\clearpage
\section{An estimate of the probability for many-body localization}
One may argue that the many-body transition occurs when the magnitude of a diagonal element becomes larger than the sum of off-diagonal elements.
Since off-diagonal elements of a Hamiltonian sampled from our local ensemble are highly correlated due to the spatial locality, the validity of this naive estimation is unclear.
However, we see below that this estimate is at least  consistent with our numerical calculation for the local random matrix ensembles, which have $M\coloneqq N d_{\mathrm{loc}}^{\, l}$ off-diagonal elements per each row.

We consider each matrix elements to be identically and independently distributed Gaussian variables.
In this case, the probability that the modulus of a diagonal element (denoted by $z$) becomes larger than that of the sum of off-diagonal elements (denoted by $x_{\alpha} +iy_{\alpha}$) in the same row is
\begin{align}
    &\quad \mathrm{Prob}\bqty{ \abs{z} \geq \abs*{ \sum_{\alpha=1}^{M} (x_{\alpha} +iy_{\alpha}) } } \nonumber \\
    &= \frac{1}{ \sqrt{2\pi} } \int_{-\infty}^{\infty} e^{ -\frac{z^2}{2} } \dd{z} 
    \frac{1}{ \sqrt{2\pi M} } \int_{-\infty}^{\infty} e^{ -\frac{X^2}{2M} } \dd{X} 
    \frac{1}{ \sqrt{2\pi M} } \int_{-\infty}^{\infty} e^{ -\frac{Y^2}{2M} } \dd{Y} \theta(z^2 -X^2-Y^2) \nonumber \\
    &= \frac{1}{ \sqrt{2\pi} } \int_{-\infty}^{\infty} e^{ -\frac{z^2}{2} } \dd{z}
    \frac{1}{M}\int_{0}^{\infty} R e^{ -\frac{R^2}{2M} } \dd{R} \theta(z^2 -R^2) \nonumber \\
    &= \frac{1}{ \sqrt{2\pi} } \int_{-\infty}^{\infty} e^{ -\frac{z^2}{2} } \dd{z}
    \qty( 1 -e^{ -\frac{z^2}{2M} } ) \nonumber \\
    &= 1 -\sqrt{ \frac{M}{M+1} } \simeq \frac{1}{2M} = \frac{1}{2N d_{\mathrm{loc}}^{\, l}},
\end{align}
where we have used the fact that the sum of $M$-Gaussian random variables with variance one is again a Gaussian random variable with variance $M$.
This estimate is too large compared with our numerical data for $\mathbb{E}[\Delta_{\infty}]$, which shows a decay as fast as an exponential in $N$.

On the other hand, the probability that the squared modulus of a diagonal element becomes larger than the sum of the squared modulus of off-diagonal elements for a random matrix with independent Gaussian matrix elements is
\begin{align}
    &\quad \mathrm{Prob}\bqty{ \abs{z}^2 \geq \sum_{\alpha=1}^{M} (x_{\alpha}^2 +y_{\alpha}^2) } \nonumber \\
    &= \frac{1}{\sqrt{2\pi}} \int_{-\infty}^{\infty} e^{ -\frac{z^2}{2} } \dd{z} 
    \prod_{\alpha=1}^{M} \qty(\frac{1}{2\pi} \int_{-\infty}^{\infty} e^{ -\frac{x_{\alpha}^2+y_{\alpha}^2}{2} } \dd{x_{\alpha}} \dd{y_{\alpha}} )^2 
    \theta( z^2 -\sum_{\alpha=1}^{M} (x_{\alpha}^2 +y_{\alpha}^2) ) \nonumber \\
    &= \frac{1}{\sqrt{2\pi}} \int_{-\infty}^{\infty} e^{ -\frac{z^2}{2} } \dd{z} \frac{ S_{2M-1} }{(2\pi)^{M}} \int_{0}^{\infty} \dd{R} R^{2M-1} e^{ -\frac{R^2}{2} } \theta(z^2 -R^2) \nonumber \\
    &= \frac{ S_{2M-1} }{(2\pi)^{M}} \int_{0}^{\infty} \dd{R} R^{2M-1} e^{ -\frac{R^2}{2} } \sqrt{\frac{2}{\pi}} \int_{R}^{\infty} e^{ -\frac{z^2}{2} } \dd{z} \nonumber \\
    &= \frac{ S_{2M-1} }{(2\pi)^{M}} \int_{0}^{\infty} \dd{R} R^{2M-1} e^{ -\frac{R^2}{2} } \qty( 1-\erf( \frac{R}{\sqrt{2}}) ),
\end{align}
where $S_{2M-1}$ is the surface area of a $(2M-1)$-dimensional unit sphere.
We estimate this integral by saddle-point approximation by assuming that $M = Nd_{\mathrm{loc}}^{\, l}$ is sufficiently large.
If we denote the integrand by $f(R)$, its derivative reads
\begin{align}
    \dv{}{R} \log f = (2M-1)\frac{1}{R} -R -\sqrt{ \frac{2}{\pi} } \frac{ e^{-R^2/2} }{ 1-\erf(R/\sqrt{2}) }. \label{Eq:DevIntegrand}
\end{align}
If $R$ is sufficiently large, we can approximate the error function as $\erf(R/\sqrt{2}) \simeq 1-\sqrt{\frac{2}{\pi}} \frac{1}{R} e^{ -R^2/2 }$.
By substituting this in Eq.~\eqref{Eq:DevIntegrand}, we obtain
\begin{align}
    \dv{}{R} \log f &\simeq (2M-1)\frac{1}{R} -R -R \simeq \frac{2(M-R^2)}{R}.
\end{align}
Therefore, $\dv{}{R} \log f (R_{\ast}) = 0 \iff R_{\ast}\simeq \sqrt{M}$, which is consistent with the approximation of the error function.
The second derivative at $R_{\ast}$ is $\dv[2]{}{R} \log f (R_{\ast}) \simeq -4\, (<0)$, which is independent of $M$.
Thus, we finally obtain the following estimate
\begin{align}
    \mathrm{Prob}\bqty{ \abs{z}^2 \geq \sum_{\alpha=1}^{M} (x_{\alpha}^2 +y_{\alpha}^2) } 
    &\simeq \frac{ 2 \pi^{M} }{(2\pi)^{M} \Gamma(M)} \sqrt{ \frac{\pi}{2} } M^{M-\frac{1}{2}} e^{ -\frac{M}{2} } \sqrt{ \frac{2}{\pi} } \frac{1}{ M^{1/2} } e^{ -\frac{M}{2} } \nonumber \\
    &\simeq \frac{1}{2^{M-1} \sqrt{2 \pi M} (M/e)^{M-1}} M^{M-1} e^{-M} \nonumber \\
    &= \sqrt{ \frac{2}{\pi e^2 M} } e^{ -M\log 2 }.
\end{align}
Since $M=Nd_{\mathrm{loc}}^{\, l}$, this probability decays exponentially with $N$, which is consistent with our result.

\clearpage
\bibliography{local} 